
\input harvmac
\def\half{{1 \over 2}}
\def\dzm{{\partial_z}}
\def\dzp{{\partial _{\bar z}}}
\def\pj{{\partial _j}}
\def\pk{{\partial _k}}
\def\pbj{{\bar\partial _{\bar j}}}
\def\pbk{{\bar\partial _{\bar k}}}

\def\ep {{\epsilon^{jk}}}
\def\epb {{\epsilon^{\bar j\bar k}}}

\def\xj {{x_j}}
\def\xk {{x_k}}
\def\xbj {{\bar x_{\bar j}}}
\def\xbk {{\bar x_{\bar k}}}
\def\pbj {{\psi^-_{\bar j}}}
\def\pbk {{\psi^-_{\bar k}}}
\def\pj {{\psi^+_j}}
\def\pk {{\psi^+_k}}
\def\a {{\alpha}}
\def\b {{\beta}}
\def\g {{\gamma}}
\def\d {{\delta}}
\def\e {{\epsilon_{\a\b\g\d}}}

\def\eb {{\epsilon_{\bar\a\bar\b\bar\g\bar\d}}}
\def\eu {{\epsilon^{\a\b\g\d}}}
\def\ad {{\dot\alpha}}
\def\bd {{\dot\beta}}

\def\t {{\theta}}
\def\tb {{\theta^*}}
\def\ta {{\theta^\alpha}}
\def\tba {{\theta^{*\ad}}}
\def\th {{\bar\theta}}
\def\N {{\nabla}}
\def\Nb {{\nabla^*}}
\def\Na {{\nabla_\alpha}}
\def\Ng {{\nabla_\g}}
\def\Nd {{\nabla_\delta}}
\def\NB {{\nabla_\beta}}
\def\Nba {{\nabla^*_\ad}}

\def\tbar {{\bar\theta}}
\Title{\vbox{\hbox{HUTP--94/A018,}}
       \vbox{\hbox{KCL-TH-94-12}}   }
{\vbox{\centerline{\bf N=4 Topological Strings}}}
\bigskip\centerline{Nathan Berkovits}
\bigskip\centerline{Mathematics Department, King's College}
\centerline{The Strand, London, WC2R 2LS, UK}
\bigskip\centerline{ Cumrun Vafa }
\bigskip\centerline{Lyman Laboratory of Physics, Harvard University}
\centerline{Cambridge, MA 02138, USA}
\vskip .2in
We show how to make a topological string theory starting from
an $N=4$ superconformal theory.  The critical dimension for this theory is
$\hat c= 2$ ($c=6$).  It is shown that superstrings
(in both the RNS and GS formulations) and critical $N=2$ strings
are special cases of this topological theory.
Applications for this new topological
theory include:
1) Proving the vanishing to all orders
of all scattering amplitudes
for the self-dual $N=2$ string with flat background,
with the exception of the three-point function and the closed-string
partition function;
2) Showing that the topological partition function of the $N=2$
string on the $K3$ background may be interpreted
as computing the superpotential in harmonic superspace generated
upon compactification of type II superstrings
from 10 to 6 dimensions;
and 3) Providing a new prescription for
calculating superstring amplitudes which
appears to be free of total-derivative ambiguities.

\Date{July 1994}
\newsec {Introduction}
Topological quantum field theories in two dimensions have become
increasingly important over the past few years.
Their first appearance in string theory
came from viewing $N=2$ topological conformal theories as special
vacua of bosonic strings which combine the ghosts with the matter
fields in a non-trivial way, and where one of the twisted
supercharges of the $N=2$ theory plays the role of the
$BRST$ operator $Q$
and the other plays the role of the anti-ghost $b$.  The critical
case for this type of `vacuum' for the bosonic string is when the
$N=2$ central charge is $\hat c =3$ \ref\wite{E. Witten, {\it Cherns-Simons
Gauge Theory as a String
Theory}, preprint IASSNS-HEP-92/45, hep-th/9207094.} .  The most
well-known examples of these theories come from twisting
sigma models on Calabi-Yau threefolds, which was studied extensively
in \ref\bcov{M. Bershadsky, S. Cecotti, H. Ooguri and C. Vafa, {\it
Kodaira-Spencer Theory of Gravity and Exact Results for Quantum
String Amplitudes},
preprint HUTP-93-A025, RIMS-946 and SISSA-142-93-EP,
hep-th/9309140\semi
M. Bershadsky, S. Cecotti, H. Ooguri and C. Vafa, Nucl. Phys. B405
(1993) 279. }.  It was found there
that the string field theory interpretation
for these theories corresponds to a theory of gravity
in three complex dimensions which was called the Kodaira-Spencer
theory of gravity.  The partition functions of this theory
have the further property that they can be viewed as computing
\bcov
\ref\nar{ I. Antoniadis, E. Gava, K.S. Narain and T.R. Taylor,
Nucl. Phys. B413 (1994) 162.}\ the superpotential terms of the four-dimensional
theory which would be obtained upon compactification
of superstrings on the corresponding Calabi-Yau.
This gave a physical reinterpretation for the meaning of $N=2$ topological
theories as isolating the $F$-term computations for standard
superstring compactifications.
Non-unitary models
of $N=2$ topological theories with $\hat c=3$ are also interesting
(one such model
is equivalent to the $c=1$ non-critical string \ref\muv{S. Mukhi and
C. Vafa, Nucl. Phys. B407 (1993) 667.}).

In a previous paper \ref\bv{N. Berkovits and C. Vafa, Mod. Phys.
Lett. A9 (1994)
653.} , we used topological
theories in a reverse way to show the
equivalence of various strings:  Roughly
speaking, we untwisted the supersymmetry
of the critical
bosonic string and found that it has a superconformal symmetry
with central charge corresponding to the critical dimension
of $N=1$ superstrings.  We then coupled it to $N=1$ supergravity
and found that the theory is equivalent to the original bosonic
string.  In this way, we managed to imbed bosonic string vacua
in a special class of vacua for $N=1$ superstrings.  Similarly,
by `untwisting' the critical $N=1$ superstring (shifting some ghost spins),
it was found to have an $N=2$ superconformal symmetry which was
then coupled to $N=2$ supergravity and shown to be equivalent
to the original $N=1$ superstring.
This gave an imbedding of $N=1$ strings into $N=2$ strings,
suggesting that $N=2$ strings have a very special
role among all the strings.  There has been further
work along these directions \ref\addwork{J. Figueroa-O'Farrill,
Phys. Lett. B321 (1994) 344\semi H. Ishikawa and M. Kato,
Mod. Phys. Lett. A9 (1994) 725\semi N. Ohta and J.L. Petersen,
Phys. Lett. B325 (1994) 67\semi N. Berkovits, M. Freeman and P. West,
Phys. Lett. B325 (1994) 63\semi F. Bastianelli, N. Ohta and J.L. Petersen,
{\it A Hierarchy of Superstrings}, preprint NBI-HE-94-20, hep-th/9403150. }.

The simplest vacuum of $N=2$ strings corresponds to string propagation
on a self-dual background in four dimensions with signature (2,2)
or with Euclidean signature (4,0)
\ref\ov{H. Ooguri and C. Vafa, Nucl. Phys. B367 (1991) 83\semi
H. Ooguri and C. Vafa, Nucl. Phys. B361 (1991) 469\semi
H. Ooguri and C. Vafa, Mod. Phys. Lett. A5 (1990) 1389.}.
 The only degree of freedom is a scalar describing the
K\"ahler class satisfying the Plebanski equation.
As was noted in \ov\ , this
theory has many features in common with what one would expect of
a topological theory.  However attempts failed to fit it into
the $N=2$ topological theory coupled to gravity since
the critical dimension for
$N=2$ topological theories is $\hat c=3$, whereas here we need a critical
dimension of $\hat c=2$.

Our main goal in this paper is to show that there is a new
class of topological theories based on the $N=4$ superconformal algebra
which includes the $N=2$ strings, thus providing
the elusive topological reformulation of $N=2$ strings.  The
critical dimension for this theory turns out to be $\hat c=2$.\foot
{Other papers \ref\anselmi{
D. Anselmi, P. Fre, L. Girardello and P. Soriani, {\it
Constrained Topological Field Theory from Twisted $N=2$
Liouville Theory}, preprint
SISSA-49-94-EP and IFUM-468-FT, hep-th/9404109\semi
D. Anselmi, P. Fre, L. Girardello and P. Soriani, {\it
Constrained Topological Field Theory}, preprint
SISSA-67-94-EP and IFUM-469-FT, hep-th/9405174.}
have recently appeared on the subject of topological
theories with $\hat c=2$, however at the present time, we do
not see a direct connection between these papers and our work.}
To view the $N=2$ strings as a special case of this string, one
starts by noting that $N=2$ matter with $\hat c=2$
(which is the critical dimension of $N=2$ strings) always has
in addition an $N=4$
superconformal symmetry.
The matter piece with $N=4$ is then twisted and the $N=2$ ghosts
are stripped off. Applying
the rules of computation which we will define
for this new $N=4$ topological theory,
one recovers
the results for the $N=2$ string amplitudes.

In a sense, this is
reversing the arrow of imbedding that we had previously found for
imbedding $N=1$ strings into $N=2$ strings.  Namely,
when applied to RNS strings imbedded into $N=2$
strings, the steps we just mentioned give back the RNS field
content untouched.  Therefore, all we have to do
is show that the rules of computation for RNS strings are
the same as the rules of the $N=4$ theory (note that RNS strings
have a twisted $N=4$ superconformal symmetry which follows
from the fact that they have an $N=2$ symmetry \bv\ and has $\hat c =2$
with integral charges), and this is easily done.   Moreover,
it will be seen that this $N=4$ reformulation
of the amplitudes gives a definite prescription
for computation which may resolve
the ambiguity question for RNS
strings.

There are further applications for this topological theory:  First
of all, it enables us to prove various vanishing theorems for the
$N=2$ strings on the flat (2,2) background.  Secondly, if we consider
$N=2$ strings on $K3$, one would expect (since
it is a topological theory and in analogy with
the Calabi-Yau threefold case) to be computing superpotential
terms for the six-dimensional theory obtained by compactifying
superstrings on $K3$.  This expectation is borne out, but
moreover the superpotential terms that one computes
{\it are automatically in the harmonic superspace} (as
would be required for supersymmetry in six dimensions).

The organization of this paper is as follows:  In section 2, we
introduce the new topological theory based on $N=4$ superconformal
symmetry.  It is shown there that
one obtains not
just a single partition function $F_g$ at each genus, but $(4g-4)+1$
of them assembled in the form of the coefficients of the homogeneous
polynomial $F_g(u_1,u_2)$ of degree $4g-4$.  The $u$'s are
related to how we choose our $N=2$ within the $N=4$ algebra
to define the string amplitudes, and are the usual
harmonic superspace vielbeins. They are parametrized by the group
$SU(2)$, and the coefficients of $F_g$ tranform as a
spin $j=2g-2$ representation of this $SU(2)$.
  We also discuss the notion
of harmonicity of the partition function for the topological
theory, which is a natural analog of the notion of holomorphicity
for $N=2$ topological theories (which was found to be anomalous \bcov).

In section 3, we show how $N=2$ strings are
equivalent to this topological theory.  We will find that
there are $(4g-4)+1$ values of $N=2$ instanton number where
the partition function does not vanish and these are in one to one
correspondence with the $(4g-4)+1$ partition functions
computed using the topological prescription. Similar results are
true for the correlation functions.

In section 4, we apply the
topological prescription to $N=2$ strings with self-dual backgrounds
in four dimensions.  We first consider the flat
$R^{2,2}$ background and show
that except for three point amplitudes and the closed-string
partition function, everything
else vanishes to all loops.
We then consider Euclidean backgrounds of the form $T^4$.
In particular, we study the $T^4$ partition function at one loop
(for special moduli corresponding to $T^4=T^2\times T^2$) and
explicitly check the harmonicity condition found in section 2.
We find that at least in this case, there is no anomaly despite
the fact that there could have been one.

In section 5, we show how the superstrings (both in the
RNS and GS formulations) fit into this new topological theory.
Since the superstring in either of these formulations can be described
by an N=2 matter sector with $\hat c=2$ (for the RNS string,
the N=2 matter sector contains both the N=1 matter fields and
the N=1 ghosts), the topological prescription
described in the previous sections can be used to calculate superstring
scattering amplitudes.
Since the topological prescription does not contain ambiguities
associated with the positions of picture-changing operators, the
resulting scattering amplitudes appear to be free of the total-derivative
ambiguities that plague conventional superstring computations.

In section 6,
we show that just as the $N=2$ topological theory on
Calabi-Yau threefolds computes superpotential terms for the
superstring compactified to four dimensions, the $N=4$
topological theory on $K3$ computes superpotential terms in
harmonic superspace for the superstring compactified to
six dimensions.
The GS method for computation
developed in \ref\ber{N. Berkovits, Nucl. Phys. B395 (1993) 77\semi
N. Berkovits, Nucl. Phys. B408 (1993) 43.}\ is shown to be an efficient means
of establishing this correspondence for both
the four-dimensional and the
six-dimensional superpotential terms.
Although the GS method is related to the
RNS method by a field-redefinition
\ref\berko{N. Berkovits,
{\it The Ten Dimensional Green Schwarz Superstring is a Twisted
Neveu Schwarz Ramond String},
preprint KCL-TH-93-12 (to appear in Nucl.Phys.B), hep-th/9308129. },
the advantage of the GS method is that
there is no need to sum over spin structures since there are no square-root
cuts and spacetime supersymmetry is manifest.

In section 7, we present our conclusions and suggestions
for further study. In this section, we also note the possibility
of constructing twisted space-time supersymmetry for self-dual
$N=2$ strings, which is particulary simple in the $N=4$ topological
framework. In appendix A, the field-redefinition
which relates the $N=2$
descriptions of the
GS and RNS superstring is reviewed. In appendix B, we show that there is
an arrow of imbeddings for topological theories. In particular,
we show that $N=2$ topological
theories (which can be viewed as special bosonic string vacua)
can be imbedded into $N=3$ topological theories (which can
be viewed as special $N=1$ superstring vacua). Unfortunately, we are
unable to construct an arrow of imbeddings that maps either of these
topological theories into the (``big'') $N=4$ topological theory.

\newsec{ A New Topological Theory}

In this section, we describe a new class of topological strings based
on $N=4$ superconformal symmetry.  There are two types of $N=4$
superconformal algebras, one ``big'' and the other ``small''.
Here we are interested in the ``small'' type which
has $SU(2)$ as its current \ref\eguch{T. Eguchi and A. Taormina,
Phys. Lett. B196 (1987) 75.}.
We will be interested in both unitary
and non-unitary realizations of this algebra\foot{In a unitary theory
with the adjoint defined by
$G^{+\dagger}=G^-$, ${\tilde G}^{+\dagger}
={\tilde G}^-$, $J^{++\dagger}=J^{--}$, and the rest of the currents
self-adjoint, the representation of the algebra has a positive-definite
norm.}
and
the most natural
unitary representations
arise from sigma models on
hyperk\"ahler manifolds.
The algebra can also be viewed as an $N=2$ algebra
$(T,G^\pm, J)$ with two additional
currents of charge $\pm 2$ , denoted by $J^{++}$ and $J^{--}$,\foot{
Their existence for the hyperk\"ahler manifolds follows from the
existence of a unique $h^{2,0}$ cohomology.}
which together with $J$ form an $SU(2)$ algebra.  Moreover under
this $SU(2)$, $G^-$ and $G^+$ generate two new supercurrents $\tilde G^+$
and $\tilde G^-$ which form pairs of doublets
$(G^+,\tilde G^-)$ and $(\tilde G^+,G^-)$.  In other words,
we have the OPE's
$$J^{--}(z)J^{++}(0)\sim {J(0)\over z}$$
$$J^{--}(z)G^+(0)\sim {\tilde G^-(0)\over z}\qquad
J^{++}(z) \tilde G^-(0) \sim {- G^+(0)\over z}$$
$$J^{++}(z)G^-(0)\sim {\tilde G^+(0)\over z}\qquad
J^{--}(z) \tilde G^+(0)\sim {- G^-(0)\over z}$$
\eqn\doubalg{J^{--}G^-(0)\sim J^{++} G^+ \sim J^{++} \tilde G^+
\sim J^{--}\tilde G^-\sim  0}
The OPE with $J$ follows from the charges of the supercurrents
which is denoted by the $\pm$ on their symbol.
Moreover, the doublets have a non-singular OPE among themselves
\eqn\comdoub{G^+ \tilde G^- \sim G^- \tilde G^+ \sim 0}
but have the following singular OPE between the elements of different
doublets
$$G^+(z) G^-(0) \sim {J(0)\over z^2} +{2T(0)+\partial J(0)\over z}$$
$$\tilde G^+(z) \tilde G^-(0) \sim {J(0)\over z^2} +{2T(0) +
\partial J(0)\over z}$$
$$G^+(z)\tilde G^+(0) \sim {J^{++}(0)\over z^2}+
{\partial J^{++}(0)\over 2z}$$
\eqn\acrdoub{G^-(z)\tilde G^-(0)\sim {J^{--}(0)\over z^2}+
{\partial J^{--}(0)\over 2z}}
The level of the $SU(2)$ current algebra is $k$ and is proportional
to the central charge $c=6k$.  For a unitary theory, $k$ is an integer
and one sees that in these cases, $c$ takes the value expected
for sigma models on a hyperk\"ahler manifold of real dimension $4k$.

We would also like to discuss how to deform the theory
in a manner that preserves the $N=4$ superconformal structure.
As is well known, \ref\dixon{L. Dixon, {\it Some Worldsheet
Properties of Superstring Compactifications, on Orbifolds and
Otherwise}, lecture given at the 1987 ICTP Summer Workshop, Trieste,
Italy, 1987.}\ to deform an $N=2$
superconformal theory, one needs chiral fields $\phi_i$
(i.e. killed by $G^+_{-1/2}$) of charge +1 and dimension 1/2,
and their conjugate $\overline \phi_i$.  One adds to the action
\eqn\deform{S\rightarrow S
+ \int t^i G^-\phi_i +\overline t^i G^+\overline \phi_i}
where here and in the following,
$G^\pm \phi$ means $\oint G^\pm(z) \phi (0)$.
Note that here and in the following, we will sometimes
concentrate only on left-movers since the equivalent argument
applies to the right-movers.
In order for this deformation to preserve the $N=4$ structure,
we will have to make sure that the deformation
is a singlet of the $SU(2)$ symmetry (since as discussed above,
the $SU(2)$ currents
and the $N=2$ algebra together generate the $N=4$ algebra).
The deformation respects the $U(1)$ part
of the $SU(2)$, so all we need to check are the invariances with
respect to $J^{++}$ and $J^{--}$.

If we look at the action of $J^{++}$
on $G^- \phi_i$, we get two terms
$$\tilde G^+ \phi_i + G^-(J^{++} \phi_i)$$
For these two terms to vanish, we thus require that
\eqn\chitil{\tilde G^+ \phi_i=0}
and
$$\oint J^{++} \phi_i=0$$
This means that $\phi_i$, being killed by the
raising operator and of charge 1, must be the top
component of an $SU(2)$ doublet.

If we look at the action of $J^{--}$ on $G^- \phi_i$, we only get
one term
$$G^-(J^{--}\phi_i)$$
and this implies that $J^{--}\phi_i$ is killed by $G^-$, i.e.,
is antichiral.  Together with the fact that the charge of $J^{--}\phi_i$
is $-1$, we see that we can expand it in terms of antichiral fields
of charge $-1$ through
some matrix $M_i^{\bar j}$:
\eqn\conv{\oint J^{--} \phi_i =M_i^{\bar j} \bar \phi_j}

We thus see that $\phi_i$ is the upper component of an
$SU(2)$ doublet with the lower component being an anti-chiral
field.
In a {\it unitary} $N=4$ superconformal
theory, a chiral field of charge +1
is automatically the top component of an $SU(2)$ doublet, with
the lower component an anti-chiral field.  However, since
we are not necessarily dealing with a unitary theory, only
chiral fields of charge +1 which satisfy this condition
can be used to deform the theory.
Conjugate statements of the above discussion of
course hold true for $\bar \phi_i$.

Therefore, we have learned that the chiral/anti-chiral
 fields of interest
are of charge $+1/-1$ and satisfy
$$G^+ \phi_i=\tilde G^+ \phi_i =0,
\qquad G^-\bar \phi_i=\tilde G^- \bar \phi_i
=0$$
\eqn\condchir{J^{--}\phi_i=M_i^{\bar j} \bar \phi_j, \qquad J^{++} \bar \phi_i
= - M_{\bar i}^j
\phi_j, \quad M_i^{\bar j}M_{\bar j}^k =\delta_i^k}
Note that these relations imply that we can use just chiral
fields (or just anti-chiral fields) to deform the action since
\eqn\chialo{G^+ {\bar \phi_i}=
M_{\bar i}^jG^+ (J^{--}\phi_j)=M_{\bar i}^j (G^+ J^{--})\phi_j
=-M_{\bar i}^j\tilde G^-
\phi_j}
It is sometimes
convenient to write the deformation of the action in the form
\eqn\dets{S\rightarrow S+\int t^i_1 G^-\phi_i - t^i_2 {\tilde G^-}\phi_i }
where in principle, $t^i_1$ and $t^i_2$ are independent.  In a unitary
theory, they are not independent and must satisfy the condition that
$$t^i_2=({\overline t^j_1}) M_{\bar j}^i \quad .$$

Suppose that we have an $N=4$ theory with trivial $\tilde G^+$ cohomology
(which often happens for non-unitary theories).  Since a chiral field
$\phi_i$
is annihilated by $\tilde G^+$ it can be solved in terms
of another field
\eqn\zepi{\phi_i= {\tilde G^+ }V_i}
where $V_i$ is $U(1)$ neutral and has
dimension 0. Since $\phi_i$ is a doublet
of $SU(2)$ and $(\tilde G^+,  G^-)$ transforms as a doublet,
there are only two possibilities for $V_i$ :  Either
it belongs to a singlet or a triplet of $SU(2)$.  Let us assume
it belongs to a singlet as will be the case in our examples.
{}From \condchir\ and the fact that $\{G^+_{-1/2},\tilde G^+_{-1/2}\}=0$,
we deduce that $ G^+ V_i$  is killed by  $\tilde G^+$.
We can thus construct out of $V_i$ a new chiral field (by using the other
 $G$-current doublet):
\eqn\const{\phi_i^{(1)}= G^+ V_i }
Note that $\phi_i^{(1)}$ is a chiral field
of charge $+1$ and dimension ${1/2}$, is
killed by $\tilde G^+$, and is a member of an $SU(2)$ doublet.
So it is a candidate to deform the theory by.  In fact, we can
repeat this process for
$\phi_i^{(1)}$ and construct a
new chiral field $\phi_i^{(2)}$.  This process can be continued
ad-infinitum and we end up with an infinite collection of
fields $\phi_i^{(n)}$ for $n>0$
which were formed from the original chiral field $\phi_i$.
Note that if the $G^+$ cohomology were also trivial,
we could have reversed this process.  For example, $\phi_i^{(1)}$
being killed by $G^+$ can be solved as $G^+ V_i$, and we can apply
$\tilde G^+ V_i$ to get a new chiral field, which is nothing
but $\phi_i$.  In this way we would get an infinite collection
of chiral fields $\phi_i^{(n)}$ for each integer $n$.
This
process is reminiscent
of the picture-changing operation of RNS which, as we
will see later, is not an accident.  Apriori, we are
not guaranteed that all these fields are different.  In fact,
we will see that the $N=2$ string with a self-dual background
gives an example where all the $\phi_i^{(n)}$ are equal to each
other up to multiplication by a $c$-number.  But in cases where
they are inequivalent, one would like to somehow identify
them as in the picture-changing operation of RNS.

Having discussed some general aspects of $N=4$ theories, we are now
ready to consider their twisting.  Since twisting involves
coupling some current to a background gauge field identified
with the
spin-connection, we have to choose a $U(1)$ in $SU(2)$ for the twist.
Once we do the twisting, the story is just the familiar
twisting of $N=2$ theories where we view an $N=4$ theory as
a special case of an $N=2$ theory. In other words,
twisting simply means that we shift
the spin content of all the fields by half their $U(1)$ charge
\ref\wit{E. Witten, Comm. Math. Phys. 117 (1988) 353\semi
E. Witten, Comm. Math. Phys. 118 (1988) 411.}\ref\eg{T. Eguchi
and S.-K. Yang, Mod. Phys. Lett. A5 (1990) 1693.}.
Thus the spin content of the fields which are affected are the following:
$$G^+,\tilde G^+ \rightarrow {\rm spin}\ 1$$
$$G^-, \tilde G^- \rightarrow {\rm spin} \ 2$$
$$J^{++} \rightarrow {\rm spin}\ 0$$
$$J^{--} \rightarrow {\rm spin}\ 2$$
Note that chiral fields $\phi_i$ of charge $+1$ and dimension $1/2$
in the twisted theory will have dimension zero.

Now we have to decide which fields in the twisted theory
we would like to identify as `physical'.  In the
case of $N=2$ topological theories, the cohomology
elements of $G^+$ form the physical spectrum.  Moreover,
any state which is $G^+$ of some other field is considered
trivial and decouples from the theory.  For the $N=4$ theories
under consideration, we have seen that the relevant chiral fields
are not only killed by $G^+$ but also by $\tilde G^+$.
Therefore the condition for a physical field should be that it is
killed by both $G^+_0$ and $\tilde G^+_0$ (note that $G$'s
have integral Fourier coefficients in the twisted theory). It is important to
note that
\eqn\comg{\{ G^+_0,\tilde G^+_0 \}=0}
 Of course
a trivial way to form a field annihilated by both $G^+$ and $\tilde G^+$
is to consider $G^+\tilde G^+ \chi$ for any $\chi$
(where we use \comg ).
This must
be the condition for topological triviality. In other words we consider
a field $\phi$ to be physical if
\eqn\physf{G^+\phi =\tilde G^+ \phi =0 \qquad \phi \sim \phi +G^+\tilde G^+
 \chi }

To complete the story, we have to discuss how to couple this
topological theory to gravity. In other words, what are the rules
for integrating over the moduli of Riemann surfaces.  In the $N=2$ twisted
theory, the story is rather simple:  $G^-$ having spin $2$
plays the role of the $b$ anti-ghost of bosonic strings, and
the twisted $N=2$ theory can be viewed as a generalization of
a bosonic string vacuum.
Before coupling to gravity, our theory can also be viewed
as an $N=2$ twisted theory.  The only new ingredient is
what we mean by a physical state, which should satisfy the condition
\physf .

If we were able to consider a {\it reduced} Hilbert space
$\tilde{\cal H}\subset {\cal H}$ where $\tilde{\cal H}$ is the
subspace of ${\cal H}$ killed by
$\tilde G^+$,
then the physical state condition \physf\
acting on this reduced Hilbert space would be
the same as the condition for an $N=2$
twisted physical field and we could use the same rules of computation.
So the question is how to
do this reduction. Using the fact that $(\tilde G^{+})^2=0$, this can
be done by simply inserting
a $\oint \tilde G^+$ around each $a$-cycle on the Riemann surface,
or in a manifestly
modular-invariant way by combining with right-movers
and integrating $\int d^2 z
 \tilde G^+ \overline {\tilde G^+}$
over the surface (when $\tilde G^+$ is holomorphic,
this surface integral reduces to integrals over the cycles in the
usual way).
We are therefore led to considering the partition function on genus
$g$ defined by the measure over moduli space ${\cal M}_g$
$$\langle |G^-(\mu_1)...G^-(\mu_{3g-3})|^2\big[\int \tilde G^+\tilde {\bar
G^+} \big]^{g}\rangle $$

However this amplitude is identically zero.  To see this, use
$\oint \tilde G^+ J=-\tilde G^+$ to replace one of the $\tilde G^+$'s
by the contour integral of $\oint \tilde G^+ $ around $J$, and pull
the contour of $\tilde G^+$ off the surface.  Since the $G^-$'s and
$\tilde G^+$'s have no singularity with it, we get zero!
There is another reason why the above formula is
not what we want.  As discussed above \physf\ , a deformation is
topologically trivial if it can be written as $G^+\tilde G^+ \chi$.
However in the above definition, it is easy to see that
adding to the action $\tilde G^+ \chi$ is already topologically
trivial since we can pull
the $\tilde G^+$ contour
off of $\chi$ and get zero by the same reasoning as above.

So instead, we will define the topological partition function to be
\eqn\defp{F_g=\int_{{\cal M}_g}\langle |G^-(\mu_1)...G^-(\mu_{3g-3})|^2\big[
\int \tilde G^+\overline {\tilde
G^+} \big]^{g-1}\int J\bar J\rangle }
which is no longer zero. Note that the contour
of $\tilde G^+$ can no longer be pulled off the surface since it hits
the $J$ and gives back $\tilde
G^+$ as the residue. For the same reason,
adding $\tilde G^+ \chi$
to the action may change the partition function.  However
if we consider adding $G^+\tilde G^+ \chi$ to the action, then
the $\tilde G^+$ contour can be pulled off of $\chi$
and converts the $J$ to a $\tilde G^+$.  Now pulling the $G^+$ contour
off of $\chi$, we encounter no residues from $\tilde G^+$. From the
$G^-$'s, we get residues which are the energy momentum tensor,
thus giving us total derivatives in the moduli which at least formally
(barring anomalies) integrate to zero.  Thus \defp\
has the correct topological symmetry.  Further
motivation for the definition \defp\ will come
in subsequent sections when we see that the
partition functions of superstrings and N=2 strings can
be viewed as special cases of it.

Before going on, it is useful for later applications to note
that we can write $F_g$ equivalently as
\eqn\deffp{F_g=\int_{{\cal M}_g}\langle |G^-(\mu_1)...G^-(\mu_{3g-4})
J^{--}(\mu_{3g-3})|^2\big[
\int \tilde G^+\overline {\tilde
G^+} \big]^{g}\rangle }
To obtain this from \defp\ , write $G^-$ as the contour of
${\tilde G}^+$ around $J^{--}$ and move $\oint \tilde G^+$ off the surface
where it only hits $J$. This leaves a residue of ${\tilde G}^+$,
which thus gives us \deffp .

Let us note that the charge violation in the genus $g$ amplitude
due to twisting is $\hat c (g-1)$.  Since the definition
\defp\ has a charge violation of $ -2(g-1)$, we learn that
the partition function is apriori non-vanishing when $\hat c=2$ (for other
values of $\hat c$, we need to insert charged operators to get a non-zero
result).
In this sense, $\hat c=2$
{\it is the critical dimension for} $N=4$ just as $\hat c=3$
is the critical dimension for $N=2$ topologically twisted theory.
Note that any
 $N=2$ theory with $\hat c=2$ and integral $U(1)$ charges
automatically has an $N=4$ symmetry \eguch\
where the spectral flow operator and its inverse, $\exp(\pm\int J)$,
provide the
raising and lowering operators of $SU(2)$.  As far as unitary
theories with a discrete spectrum, the only examples are
based on sigma models on $T^4$ and on $K3$.

So far, we have avoided an important point. We have discussed how
$N=2$ superconformal symmetry
can be realized in an $N=4$ algebra and have
used  the $N=2$ to discuss twisting and to define
the topological amplitude. However we should also ask how many
ways one can view an $N=4$ theory as an $N=2$ theory.
If there is more than one way, as we will find, then there
is more than one topological theory associated to a given $N=4$ theory.

To pick an $N=2$ in an $N=4 $ theory, we have to do two things:
First we have to choose the $U(1)$ current
$J$ of $N=2$ in the $SU(2)$ of the
$N=4$.  Secondly, we have to decide which currents are $G^+$ and $G^-$.
In fact, there is an $SU(2)$ worth of consistent choices for $G^\pm$.
To see this, note that we have two doublets of $G$'s and
the $N=4$ algebra is unmodified if we rotate
them with each other by an $SU(2)$.  To avoid confusion with the $SU(2)$
that is part of the current algebra of $N=4$, we will call this the
{\it flavor} $SU(2)_f$.  The $SU(2)$ which is part of the $N=4$
algebra we shall call the {\it color} $SU(2)_c$.
We write the flavor rotation as
$$\widehat{{\tilde G}^+}(u)=u_1 {\tilde G}^+ +u_2 { G^+}$$
$$\widehat {G^-}(u)=u_1 { G}^--u_2{\tilde G^-}$$
$$\widehat {{\tilde G}^-}(u)=u_2^*{\tilde G}^- - u_1^*{ G^-}$$
\eqn\rot{\widehat {G^+}(u) =u_2^*{G}^+ +u_1^* {\tilde G^+}}
where
$$|u_1|^2+|u_2|^2=1$$
Note that we are using the harmonic notation of \ref\harmr{A. Galperin,
E. Ivanov, S. Kalitzin, V. Ogievetsky and E. Sokatchev, Class. Quant. Grav. 1
(1984) 469.}\
in which the complex conjugate of $u_a$ is $\epsilon^{ab} u^*_b$
(i.e. $\overline {(u_1)}=u^*_2$ and $\overline{(u_2)}=-u^*_1$
where $*^2=-1$).
Furthermore, the SU(2) indices on
$u$ and $u^*$ can be raised with the antisymmetric tensor
$\epsilon^{ab}$.
For example, the normalization condition on the $u$'s can be written
as $u^a u^*_a =1$.

Now let us see how different embeddings of the $N=2$
algebra in $N=4$ modify the twistings and amplitudes.  As far
as twisting, note that $SU(2)_c$ is an internal symmetry of the theory
which rotates different choices of $U(1)$ into one another.
Since the choice of twisting involves the choice for this $U(1)$,
different twistings can be obtained by rotating the whole field space
using $SU(2)_c$.  In particular, any computation we do with any given
choice is equivalent to a conjugate computation done with the conjugated
$U(1)$.  Thus the different choices of $U(1)\subset SU(2)_c$
do not lead to any new theories.  In particular, this means that
any computations done with
the anti-topological twisting (where $G^+$ is dimension 2 and
$G^-$ is dimension 1) is equivalent to a computation with
the topological twisting where the rotation in field space sends
$$G^+\to {\tilde G}^-, \quad
G^-\to {\tilde G}^+$$
\eqn\equcol{\tilde G^+\to -G^-,\quad\tilde G^- \to -G^+}

The $SU(2)_f$ is more interesting as far as getting a new
theory.  This is because $SU(2)_f$ is not realized by a symmetry
(there is no current associated with it) and
 the theory we get with the rotated $G$'s \rot\
in general results in a new theory.
The definition of the flavor-rotated amplitude is
given by \defp\ (or \deffp ), but now we use
$(\widehat{{\tilde G}^+}(u),\widehat{ G^-}(u))$,
instead of $(\tilde G^+,G^-)$
and we end up with $F_g(u_1,u_2)$.  So the topological partition function
is a homogeneous polynomial in $u_1,u_2$ of degree $4g-4$:
\eqn\homp{F_g(u_1,u_2)=\sum_{n=-2g+2}^{2g-2}{(4g-4)!\over (2g-2+n)!(2g-2-n)!}
F_g^n u_1^{2g-2+n} u_2^{2g-2-n}}
$$=\int_{{\cal M}_g}\langle |\widehat{G^-}(\mu_1)...\widehat{G^-}
(\mu_{3g-3})|^2\big[
\int \widehat{\tilde G^+}\widehat{\overline {\tilde
G^+}} \big]^{g-1}\int J\bar J\rangle $$

Note that $u_1$ and $u_2$ transform as a doublet under $SU(2)_f$ where
\eqn\flavorrot{J^{+}_f =u_1 {d\over du_2}+u^*_1 {d\over du^*_2},\quad
J^{-}_f= u_2 {d\over du_1}+u^*_2 {d\over du^*_1}, }
$$J^3_f=\half(u_1 {d\over du_1}
-u_2 {d\over du_2}+u^*_1 {d\over du^*_1}-u^*_2 {d\over du^*_2}).$$
So the $F_g^n$'s form a spin $2g-2$ dimensional
representation of $SU(2)_f$ where $n$ labels the $J_3$ eigenvalue
of $SU(2)_f$.  Thus taking the different imbeddings of the
$N=2$ in $N=4$ theory would lead us not to a partition function, but to
a partition vector $F_g^n$.  Knowing $F_g^n$ for all $n$
allows us to compute the amplitudes for an arbitrary imbedding
of $N=2$ in a given $N=4$ theory, where the imbedding
is parametrized by the $u_i$'s.
The reason for the combinatorial factor in the definition of $F_g^n$ above
is that for a given
power of $u_1$ and $u_2$, there are as many correlation functions
to compute as indicated by the combinatorial factors.  We will see
below that, barring anomalies, all of these correlation functions are equal
to $F_g^n$.
Note also that we can do the left-moving and right-moving
$SU(2)_f$ rotations seperately which would lead us to introducing
another set of independent $u$'s for the right-movers. In particular,
$F_g$ would be viewed as a $(2g-2,2g-2)$ represention of
the two $SU(2)_f$'s.  As we have done in \homp , we will continue
suppressing the right-moving $u$'s to avoid excessive notation.

It will sometimes be convenient to collect all the different $F_g^n$'s
into a single function by introducing a periodic variable $\theta$ and
define\foot{As we will see later in the context of $N=2$ strings,
$\theta$ can be viewed as the $\theta$-angle of the the U(1) gauge
symmetry of the $N=2$ strings.}
$$F_g(\theta)=\sum_{n=2-2g}^{2g-2}{\rm exp}(i n \theta)\ F_g^n.$$
This new parameter $\theta$ may be viewed as an extra ``coupling constant''
for the $N=4$ topological strings. In this way, we get $SU(2)/U(1)$
inequivalent $N=4$ topological theories, where the $U(1)$ is modded
out because it can be undone by a shift in $\theta$.

Let us show  one basic property of $F_g^n$ which also
illustrates an application of $SU(2)_c$ symmetry.
Consider the $SU(2)_c$ rotation
which takes $J$ to $-J$ and transforms the $G$'s as in
\equcol.  Then conjugating the topological
theory with this symmetry changes the topological twisting to
the anti-topological twisting.  Since this is a symmetry
operation, it does not affect the topological amplitudes.
We thus learn that
\eqn\toant{F_g^{n,{\rm top.}}=F_g^{-n,{\rm anti-top}}}
where $F_g^{\rm anti-top}$ is the amplitude defined in \homp\ but
with the anti-topological twisting where
$G^+\leftrightarrow G^-$ and ${\tilde G^+}
\leftrightarrow {\tilde G^-}$.
For a {\it unitary} $N=4$ theory,
the anti-topological theory is the complex cojugate
of the topological theory implying
that
$$F_g^{n}=\overline{F_g^{-n}}$$

So far, we have talked about the amplitude for genus $g>1$.
We have to note two special cases:  genus 0 and genus 1.
Genus 0 by itself has zero partition function by $SL(2,C)$
invariance, so we will discuss it when we talk about
correlation functions.  For $g=1$, the definition is a simple generalization
of the $N=2$ partition function
 \ref\cv{S. Cecotti and C. Vafa, Comm. Math. Phys. 157
(1993) 139.}\bcov .  The main point
is that we have an extra $\int J\overline J$.  In other words, we have\foot{
In general, the zero point function at genus one may be ill-defined
if there is a moduli-independent infinity
due to the ground state contributions which needs to be deleted.
For this reason, it is better
to motivate the above definition of $F_1$ by noting that its
derivatives with respect to the moduli generate the one point functions
on the torus.}
\eqn\eff{F_1=\int d^2\tau \ {\rm Tr}(-1)^{\rm F}
F_L^2 F_R^2 q^{H_L}{\bar q}^{ H_R}}
where $F_L=\oint J$ and $F_R=\oint {\bar J}$.
If we take the derivative of $F_1$ with respect to the moduli $t^i$,
we replace a single $F_L F_R$ in \eff\
with $|G^-|^2$ and an insertion of the
chiral field $\phi_i$. Since we have an extra $F_L F_R$ (which comes from
$\int J\bar J$ upon using the Riemann identities), this is the natural
generalization of \defp\ to genus 1.
Note that if we had only the first power in $F_L$ and $F_R$ as in the
$N=2$ case, then \eff\ would have vanished for $N=4$ theories by $SU(2)$
symmetry, since $F_L$ is
mapped to $-F_L$ by a rotation in $SU(2)_c$.  Instead
$F_L^2$ is proportional to the casimir when summed over all states
of a given representation, and is not apriori zero.

We will now discuss the correlation functions for the topological theory.
As discussed above, we need chiral fields satisfying \condchir .
Note that the physical field condition of \condchir\ is true
no matter what $u$'s we choose.  We
thus consider
\eqn\corfu{F_{g,i_1...i_n}(u_1,u_2)=\int_{{\cal M}_{g,n}}
\langle \big| \prod_{j=1}^{3g-3+n} \widehat {G^-}(\mu_i)\big|^2 \big[\int
\widehat{{\tilde G}^+} \widehat{\overline {\tilde G^+}}\big]^{g-1}\int J\bar J
\quad \phi_{i_1}...\phi_{i_n}
\rangle }
where chiral fields
$\phi_i$ are inserted at arbitrary points (recall that they have
dimension zero in the
twisted theory), and we have chosen to treat
the moduli of punctures and the Riemann surface symmetrically.
Note that $F_{g,i_1...i_n}(u_1,u_2)$ is a homogeneous polynomial in $u_i$ of
degree $4g-4+n$ whose coefficients transform as a
spin $2g-2+{n\over 2}$ representation of $SU(2)_f$.

Let us now discuss the amplitudes on the sphere.  We will concentrate
on the critical case with $\hat c=2$.
  The first non-vanishing
case to consider, by $SL(2,C)$ invariance, is the three-point function.
We need three fields each of dimension zero, but with
a total charge of $+2$ to cancel the background charge of $-2$.
Since the chiral fields  have dimension 0 and charge 1 in the twisted
theory, three of them would give zero
by charge conservation. This is similar to the vanishing we would
have gotten at higher genus had we inserted $g$ $\tilde G^+$'s
around the cycles.  We solved this problem
by deleting one of the ${\tilde G^+}$'s.
We can do a similar thing here if the ${\tilde G^+}$ cohomology
is trivial, as will turn out to be the case in many applications.
As noted above \zepi\ ,we can then write the chiral primary field $\phi_i$
in terms of
$$\phi_i={\tilde G^+}V_i$$
where $V_i$ has charge 0 and dimension 0.
The three point amplitude can then be defined as
\eqn\thtre{C_{ijk}=\langle \phi_i \phi_j V_k\rangle}
Note that despite its appearance, $C_{ijk}$ is symmetric in its
indices.  To see symmetry in the indices $j$ and $k$, write
$\phi_j={\tilde G^+}V_j$ and pull the ${\tilde G^+}$ contour
off the surface. It only gives a contribution when it passes
through $V_k$, converting it to $\phi_k$ and thereby establishing
the symmetry.
Similarly for $n>3$, we define the tree amplitude to be
\eqn\trnpt{C_{i_1i_2...i_n}=\int_{{\cal M}_0,n}\langle G^-(\mu_1)
...G^-(\mu_{n-3})\phi_{i_1}...\phi_{i_{n-1}} V_n \rangle}
which is symmetric in its indices for the same reason.

In defining the above amplitude, we had to assume that the ${\tilde G}^+$
cohomology is trivial.  Even if this is not the case, we can still
define $n>3$ point amplitudes on the sphere by noting that for
$n>3$, we can write $V_n$ in the integrated form $G^-V_n$.
Using $[J^{--},\tilde G^+]=G^-$,
the definition of ${\tilde G}^+V_n=\phi_n$, and the fact that
$J^{--}$ annihilates $V_n$,
we have
$$G^-V_n=J^{--}\phi_n$$
Thus we can define
\eqn\btre{C_{i_1...i_n}=\int_{{\cal M}_0,n}\langle G^-(\mu_1)
...G^-(\mu_{n-4}) \phi_{i_1} ...\phi_{i_{n-1}} \int J^{--}\phi_{i_n}
\rangle}
The symmetry in indices follows from conformal invariance, which
allows putting $J^{--}$ at other $\phi_i$.  This definition
of the genus zero correlation function has the advantage of making
sense whether or not the ${\tilde G}^+$ cohomology is trivial.
Moreover, it is the natural generalization of \deffp\ to genus 0.
 However note that in general, we can only define the three point
function when ${\tilde G}^+$ has trivial cohomology.

We can once again introduce the $u_i$'s, in which case
\btre\ will be a homogeneous polynomial of degree $n-4$
for $n\geq 4$.  This fits well with the higher genus correlation function
which is a homogeneous polynomial of degree $4g-4+n$ in the $u_i$'s.

Let us now discuss the relation of the correlation
functions \corfu\ to the partition function $F_g$ where
the action has been deformed as in \dets .
The relation is
\eqn\recor{F_{g,i_1...i_n}(u_1,u_2)=
(u_a {D\over Dt_a^{i_1}})...(u_a {D\over Dt_a^{i_n}})F_g(u_1,u_2)}
where $D$ denotes an appropriate covariant derivative (see
\bcov\ for the definition of $D$ and for the reason
we need {\it covariant} derivatives).
To get \recor\ , one writes \corfu\ in the form
where the fields are put in the $(1,1)$ picture $\widehat{ G^-}
 \phi_i$ by taking
the appropriate beltrami differentials.

For $N=2$ topologically twisted theories, there
is a notion of {\it holomorphicity}. Namely one formally expects
the partition function to be independent of the anti-chiral
deformation, i.e. $\bar \partial_i F_g=0$, but there are
anomalies which can be understood \bcov .
It is natural to discuss the analogous notion for
the topological amplitude in the $N=4$ topological theory.
Note that the anti-chiral fields no longer decouple
since $G^+ {\overline \phi_i}$ is not a trivial field in the new definition
of the topological theory.  This is also clear given the
fact that we can write it as some combination of
$\tilde G^- \phi_j$'s from \chialo .

However in the $N=4$ topological theory,
perturbation by $G^+\tilde G^+ \chi$ is expected to be trivial
up to total derivatives, which could in principle lead
 to anomalies due to boundary contributions.  For this
perturbation to be neutral (and thus not trivially zero),
$\chi$ would have to be a charged field of dimension 1 and charge $-2$
in the twisted theory, or in the untwisted
theory, it would correspond to a field of dimension
0 and charge $-2$.  Such a $\chi$ does not exist in unitary theories as
it would violate unitarity bounds (in
unitary theories, the dimension is always greater than or equal to
half the absolute
value of the charge).  However we will
be interested in non-unitary theories as well, and in
such cases there could exist such $\chi$'s and
we could end up with such anomalies which deserve to be
studied (note that the non-unitary case may be trickier
than the one encountered for unitary theories in \bcov\
because of potential contributions from negative energy states at the
boundaries of moduli space).

Nevertheless, there is another notion of `holomorphicity' which is
applicable even to unitary theories.  One can formally argue that
\eqn\harmo{\epsilon_{ab}{d\over du_a}{D\over Dt^i_b}F_g(u_1,u_2)=0,
\qquad \epsilon_{ab}{d\over du_a}{D\over D{t^{*i}_b}}F_g(u_1,u_2)=0}
where the first equation is evaluated using the deformed action
$$S+ \int t_1^i G^-\phi_i -t_2^i\tilde G^- \phi_i$$
and the second equation is evaluated using the deformed action
$$S+ \int t_2^{*i} G^+\overline\phi_i +
t_1^{*i} \tilde G^+\overline \phi_i.$$
We will sometimes refer to \harmo\ as the ``harmonicity'' condition.
The two equations
are consequences of one another since from \chialo ,
$M_{\bar i}^{j} t_1^{*i}=t_1^j$ and
$M_{\bar i}^{j} t_2^{*i}=t_2^j$ (we are using the
same ``harmonic'' notation for the $t$'s as we did for the
$u$'s in \rot\ ).  It will be more convenient
to prove the second equation involving $t^*$'s.

To prove this equation, it is convenient to expand $F_g$ in powers of $u$'s.
We can then rewrite the above equation as the statement that
inserting
the operator $\int G^+\bar \phi_i$ in $F^n_g$
is equivalent to inserting
the operator $\int {\tilde
G^+}\bar \phi_i$
for $F_g^{n-1}$ (as usual, we are ignoring the right-movers in
our discussion to reduce the notation) .   Let us see why
this is true modulo total derivatives in moduli.

{}From $F_g^n$, we get $(2g-2+n)!(2g-2-n)!/(4g-4)!$ times a sum of
contributions of the form
$$\langle G^-(\mu_1)...G^-(\mu_r){\tilde G^- }(\mu_{r+1})...{\tilde G^-
}(\mu_{3g-3})$$
$${\tilde G^+}(z_1)...{\tilde G^+}(z_s)G^+
(z_{s+1})...G^+(z_{g-1})J(z_g)\rangle$$
where $r+s=2g-2+n$. In other words $2g-2+n$ is the number of
one type of doublet, $2g-2-n$ is the number of the other type, and
all allowed values of $r$ and $s$ subject to the above condition occur
as part of the definition of $F_g^n$.
We will now see that
up to total derivatives in moduli (which we are going
to ignore for the moment), all different allowed
values of $r,s$ give the same result. Since there are
$(4g-4)!/(2g-2+n)! (2g-2-n)!$ such terms in $F_g^n$, $F_g^n$ is normalized
to be equal to any one of these terms (up to total derivatives).

To see that all allowed
values of $r,s$ give the same result, write one of the
$G^-$'s (to simplify the notation, say the first one)
as the contour of ${\tilde G^+}$ (which is a current of dimension 1)
around the current $J^{--}$, and contour
deform ${\tilde G^+}$.  When the contour hits $\tilde G^-$'s,
we get residues of the energy momentum tensor
which lead to total derivatives that we are ignoring.
Otherwise, it has no other residues except with $J$ which gives us
$$\langle J^{--}(\mu_1)...G^-(\mu_r){\tilde G^- }(\mu_{r+1})...{\tilde G^-
}(\mu_{3g-3})$$
$${\tilde G^+}(z_1)...{\tilde G^+}(z_s)G^+ (z_{s+1})...G^+(z_{g-1}){\tilde
G^+}(z_g)\rangle$$

Now we can write one of the $G^+$'s (say the last one) as the contour
of $G^+$ around $J$.  Contour deforming the $G^+$ we see that, again
up to total derivatives,
it only picks up a residue when it hits $J^{--}$ which gives us
$$\langle{\tilde G^-}(\mu_1)...G^-(\mu_r){\tilde G^- }(\mu_{r+1})...{\tilde
G^- }(\mu_{3g-3})$$
$${\tilde G^+}(z_1)...{\tilde G^+}(z_s)G^+ (z_{s+1})...J(z_{g-1}){\tilde
G^+}(z_g)\rangle$$
We thus see that we have changed $r\rightarrow r-1$ and $s\rightarrow
s+1$, which is what we wished to show.

Now we consider inserting the operator $G^+ \overline \phi_i$ in $F_g^n$.
By the above argument, all we have to show is that we can convert it to
${\tilde G^+} \overline \phi_i$, while at the same time changing
the number $r+s$ by $-1$.
By contour deforming $G^+$ off of $\overline \phi_i$,
we can put $G^+$ on the $J$ using the above procedure.
Next, we pull one of the
$\tilde G^+$'s (by writing it as the contour of $\tilde G^+$ around $J$)
off the surface, which hits only $\phi_i$ to give us $\tilde G^+
\overline \phi_i$.
We have therefore established \harmo\ up to total derivatives.

Of course, it is still possible that
\harmo\ has anomalies due to boundary contributions, a
phenomenon which does take place in the $N=2$ topological
theory coupled to gravity.  It would be rather interesting
to carefully study the potential existence of similar
anomalies in this case.

Let us consider the genus 1 generalization of \harmo.
In this case, since the partition function has no $u$ dependence,
we should replace $F_1$ in \harmo\ with the one point
function on a torus, $F_{1,i}(u_1,u_2)$.
Using definition \recor\ for $F_{1,i}$, we therefore need to prove up
to total derivatives that
\eqn\gina{\epsilon_{ab}{d\over du_a} {d\over d{t^{*j}_b}} F_{1,i}=
\epsilon_{ab}{d\over du_a} {d\over d{t^{*j}_b}}
(u_c {d\over d t^i_c}) F_{1}=
\epsilon_{ab}{d\over d t^{*j}_b}{d\over dt^i_a}F_1=0}
for all $i$ and $j$. We shall be rather brief in our proof as the arguments
are similar to the above.

Start with $\partial_{t^{i}_1}F_1$ which involves
$$G^-(z_1) J(z_2) \phi_i(z_3)$$
and consider the $\partial_{t^{*j}_2}$ derivative of it, which gives
$$-G^-(z_1) G^+(z_2) \phi_i(z_3) \overline{\phi_j }(z_4)$$
Now we write $G^-$ as a contour integral of ${\tilde G^+}$
around $J^{--}$ and pull the contour off the surface,
leaving only the contribution from the residue at ${\overline \phi_j}$
which is
$$J^{--}(z_1)G^+(z_2) \phi_i(z_3) \tilde G^+
 {\overline \phi_j}(z_4)$$
Finally, writing $G^+$ as the
contour of $G^+$ around $J$ and
 pulling the $G^+$ contour off the surface
(and noting there is no contribution from the ${\overline \phi_j}$
terms due to \chialo), we are left
with
$$-\tilde G^-(z_1)J(z_2) \phi_i(z_3) \tilde G^+
 {\overline \phi_j}(z_4)$$
which is just the definition of
$\partial_{t^{*j}_1}\partial_{t^i_2}F_1$ and so we are done.
Note that equation \gina\ can also be rewritten
using \chialo\ as
\eqn\oneloop{\partial_i {\overline \partial_{\bar j}}
F_1=M_i^{\bar i}M_{\bar j}^j
{\overline \partial_{\bar i}}{ \partial_j}F_1}
where holomorphic derivatives refer to the first component of $t$ and
anti-holomophic derivatives refer to the second
component of $t^*$.
Again, there was room for total derivative terms contributing
to \oneloop\ and we have not considered them. However
we shall see in a later section that in the context
of a special example, the potential anomalies do not appear.
It would be interesting to study potential
holomorphic anomalies more carefully and see if they  are
always absent.

Before closing this section, we will show that when the theory is unitary
and all chiral fields are primary,
the topological correlation functions can be expressed in
a slightly different form which will be useful for some
later applications.  Consider \corfu\ and study
its dependence on the position of the $\widehat{\tilde G^+}$'s before
integration over the Riemann surface.  Since $\widehat{\tilde G^+}$ is a
holomorphic current, the dependence on its position is going
to be meromorphic.  Since the chiral fields are primary,
$\widehat{\tilde G^+}$ has no singularities at the $\phi_i$'s.
Furthermore, since the theory is unitary,
$\widehat{\tilde G^+}$ has no unphysical poles coming
from the negative-energy fields. Therefore, the only potential singularity
as a function of its position could come from the pole at
$J^{--}$.  But the amplitude for the residue of this pole,
which involves replacing $J^{--}$ with $\widehat{G^-}$, is zero because
we can now write any of the $\widehat{\tilde G^+}$'s as a contour
of $\widehat{\tilde G^+}$ around $J$ and pull the contour off the
surface.

Thus we see that
as a function of the $g$ positions, $v_i$, of the ${\tilde G^+}$'s,
we have a holomorphic, totally anti-symmetric object which
transforms as one forms.  This implies that the dependence of the
correlations on $v_i$ is given by ${\rm det}{\omega}^i(v_j)$
where $\omega^i$ are the holomorphic 1-forms.  Multiplying
and dividing by this factor and integrating over the positions of $v_i$, we
learn that
$$F_g={\rm det}\ {\rm Im} \tau
\int_{{\cal M}_g}
\big|{\rm det}{\omega}^i(v_j)\big|^{-2}\cdot$$
\eqn\unin{\langle \big|G^-(\mu_1)...J^{--}
(\mu_{3g-3+N}) \big|^2 \prod_{i=1}^g \big|{G^+(v_i)}\big|^2\phi_1...\phi_N
\rangle}
where ${\rm det}\ {\rm Im} \tau$ arises from $\int |{\rm det}
{\omega}^i(z_j)|^2$, $\tau$ is the period matrix, and the $v_i$'s can be
any $g$ points on the surface.

 \newsec{$N=2$ String as an $N=4$ Topological String}
We have seen that the critical $N=4$ topological string
has ${\hat c}=2$.  We have also noted that every $N=2$ superconformal
theory with ${\hat c}=2$ (and with integral $U(1)$ charges) automatically
gives rise to a critical $N=4$ superconformal theory.  Given the
fact that every $N=2$ superconformal theory with $\hat c=2$ can be
viewed as a background for $N=2$ strings, it is natural to
ask what is the relation between these two string theories.
  In this section, we will show
that these two strings are indeed equivalent.

Before going into the detail of the proof of this equivalence,
let us outline some aspects of the proof
which may be helpful in following the arguments.  $N=2$ strings
require integrating over $N=2$ supergeometries which
in genus $g$ has the usual
 $3g-3$ bosonic moduli, $4g-4$ fermionic moduli,
and $g$ additional bosonic moduli correponding to integrating
over $U(1)$ flat connections.  Moreover, we have to sum over all
possible $U(1)$ instanton numbers. It turns out that the only
non-vanishing partition functions have
 instanton numbers $n_I$ satisfying $2-2g\leq n_I\leq 2g-2$.
The rest vanish due to superghost zero modes.

We will show that
\eqn\inseq{F_g^n=A_g^n}
where $F_g^n$ is the topological partition function
defined in \homp\ and $A_g^n$ is the N=2 string partition function
on a surface of genus $g$ and instanton number $n$
(note that the instanton number
is identified with the eigenvalue of $J_3$ flavor where $u_1$
has charge +1 and $u_2$ has charge $-1$).\foot{
In the case of $N=2$ strings (for a fixed realization of the $N=2$ algebra),
the complete answer for the partition function at genus $g$
is given by
$$A_g(\theta)=\sum_{n_I} {\rm exp}(i n_I \theta )A_g^{n_I}$$
where $\theta$ is the instanton $\theta$-angle
for the $U(1)$ field of the $N=2$ string.
This suggests that in the $N=4$ topological formulation, we should also
consider the corresponding sum as the definition of the genus $g$ amplitude,
where we think of $\theta$ as an additional coupling constant.
This means that there are not an $SU(2)_f$ worth of different theories
associated to a given $N=4$ superconformal theory, but instead,
an $SU(2)_f/U(1)$ worth of different theories
where $\theta$ labels the $U(1)\subset SU(2)_f$.}

Note that for a given
$N=2$ vacuum, we have an $SU(2)$ worth of inequivalent ways of defining
its $N=2$ fermionic generators, just as in the  topological case. Labeling
the different choices with the harmonic variables $u_i$, we will show that
\eqn\trans{
F_g(u_1,u_2)=\widehat A_g^{2g-2} (u_1,u_2)}
where $\widehat A_g^{2g-2}(u_1,u_2)$ is the N=2 string partition
function at instanton-number
$2g-2$ when the N=2 BRST operator is constructed using $\widehat G^+$
and $\widehat G^-$ as the matter part of the two fermionic generators
(although $\widehat G^+$ depends on $u^*$, it will be
found that $\widehat A_g^{2g-2}$ is actually independent of $u^*$).

The first step in the equivalence
proof will be to show that \inseq\
holds when $n=2g-2$, i.e. $F_g^{2g-2}=A_g^{2g-2}$.
After checking that the ghost partition function cancels out (with a
 judicious choice of where the various
operators are inserted), we will be left with a twisted matter
theory and some additional operators which reproduces \deffp .
The second step will be to show
 that for a different realization of $N=2$ (parametrized
by $u$), $\widehat
A_g^{2g-2} (u_1,u_2)$ is a polynomial in $u$ where the coefficients
are the $A_g^n$'s  calculated using the original unrotated
 $N=2$ superconformal generators.
By comparing with the dependence of  $F_g(u_1,u_2)$ on the $u$'s in
\homp\ , this
proves \inseq\ for all values
of $n$.  We now proceed with constructing the proof in detail.

The $N=2$ string prescription for calculating the partition function
on a genus $g$ surface of instanton number $n_I$ is given by :

$$A_g^{n_I}=\prod_{i=1}^g\int  d^2 M_i \prod_{j=1}^{3g-3}\int  d^2 m_j$$
\eqn\given{
<| (\int_{a_i} \tilde b)
(\int \mu_j b))
(Z^-)^{2g-2+n_I}
(Z^+)^{2g-2-n_I}I^{n_I}\tilde c(x_0)|^2>}
where $M_i$ are the U(1) moduli which take values in
the Jacobian variety
${\cal C}^g/({\cal Z}^g+\tau {\cal Z}^g)$,
$m_j$ are the Teichmuller parameters,
$\int_{ a_i}$ is an integration around the $i^{th}$ $a$-cycle,
$\mu_j$ are the beltrami differentials for the Teichmuller parameters,
$(\tilde b, \tilde c)$
are the U(1) ghosts (the $\tilde c$ ghost can
be inserted anywhere on the surface since
only its zero mode contributes), the N=2 superconformal ghosts
are bosonized as $(\gamma^+,\beta^-)=(\eta^+ e^{\phi^+},
\dzm\xi^- e^{-\phi^+})$
and $(\gamma^-,\beta^+)=(\eta^- e^{\phi^-},
\dzm\xi^+ e^{-\phi^-})$, the picture-changing operators are
$$Z^-=\{Q,\xi^-\}= e^{\phi^+}
[G^-+ (b-\half\dzm \tilde b)\eta^- e^{\phi^-}
 -\tilde b\dzm\eta^- e^{\phi^-}]+c\dzm\xi^-,
$$
$$Z^+=\{Q,\xi^+\}= e^{\phi^-}
[G^+ + (b+\half\dzm \tilde b)\eta^+ e^{\phi^+}
+\tilde b\dzm\eta^+ e^{\phi^+}]+c\dzm\xi^+,$$
$I=\exp(\int J_{total})=\exp(\phi^- -\phi^+ +\int J +c \tilde b)$
is the BRST invariant
instanton-number-changing operator (or spectral flow
operator), and $n_I$ is the instanton
number of the U(1) gauge field.  Note that the relative number of
picture-changing
operators depends on the instanton number because the number
of zero modes of $\beta^\pm, \gamma^\pm$ (which carry U(1) charge)
depend on the instanton number.

The above amplitude vanishes for
$|n_I|>2g-2$ since for $n_I>2g-2$, there are zero modes
of $\gamma^-$ which can not be absorbed (this is similar to
the vanishing of tree amplitudes with less than three
external vertex operators
because of the zero modes of the $c$ ghost). Note that there are no
inverse-picture-changing operators in the N=2 string which could absorb
the $\gamma^+$ zero modes. Furthermore,
note that the $c \tilde b$ term in $I$ can
be ignored in calculations since there are no extra $\tilde c$'s available
to absorb the $\tilde b$.

In order to end up with the topological prescription, we will first
 choose
$n_I=2g-2$ and  twist the N=2 algebra so that the conformal weight
of all fields is shifted by half their U(1) charge (e.g., $G^-$ now
has conformal weight 2 and $G^+$ has conformal weight 1).
Twisting the algebra is equivalent to
removing $g-1$ $I$'s from the above expression for the scattering amplitude
(note that the locations of the $I$'s are irrelevant since moving
$I$ is equivalent to shifting the U(1) moduli).
Furthermore, the locations of the picture-changing operators should
be chosen so that the path integrals over the N=2 ghosts cancel out. Note
that when $n_I=2g-2$,
the integrand of $A_g^{n_I}$ is independent of the positions of the $Z^-$'s
since
$\dzm Z^- = \{Q, \dzm \xi^- \}$ and there are no $\eta^+$'s available to
absorb the $\xi^-$. This is a special feature that is only true for
$n_I=2g-2$ since otherwise, $Z^+$'s are present which contain
$\eta^+$'s.

A convenient choice for the locations of the picture-changing operators
is to sew in
$3g-4$ of the $Z^-$'s using the first $3g-4$ beltrami differentials
$\mu_j$'s,
and to sew
$I^{g-1}$ with the last beltrami differential (for notational convenience,
we shall assume that this beltrami differential is concentrated at the
point $w$).
The remaining $g$ $Z^-$'s should be located at
points $v_1,..., v_g$ satisfying  $[v_1 + ...+ v_g -g w]_i =M_i$
where the divisor $[x-y]$ is defined as
$[x-y]_i=  \int_x^y  \omega_i$ and $\omega_i$ are the $g$ abelian one-forms
satisfying $\int_{a_j} \omega_i =\delta_{ij}$
and $\int_{b_j} \omega_i=\tau_{ij}$.

In this special picture, the above scattering amplitude takes the form:
$$A_g^{2g-2}=\prod_{i=1}^g \int d^2 M_i \prod_{j=1}^{3g-3}\int  d^2 m_j$$
$$
<| (\int_{a_i} \tilde b)
\prod_{k=1}^{3g-4}(\int\mu_k b)
\delta(\int\mu_k \beta^-)(\int\mu_k G^-)
\mu_{3g-3}(w)b(w) I^{g-1}(w)$$
$$ Z^-(v_1)...Z^-(v_g)  \tilde c(x_0)|^2>.$$

The next step is to change variables from $M_i$ to $v_i$ using the
Jacobian factor $\det (\omega_k(v_i))$. Since the integrand evaluated at $M_i=
[v_1 + ...+ v_g -g w]_i$ is equivalent to inserting
$I(v_1)...I(v_g) I^{-g}(w)$ and evaluating at $M_i=0$,
the above amplitude can be written as:
$$A_g^{2g-2}=\prod_{i=1}^g\int
d^2 v_i |\det(\omega_k(v_i))|^2 \prod_{j=1}^{3g-3}\int d^2 m_j$$
$$<| (\int_{a_i} \tilde b)
\prod_{k=1}^{3g-4}(\int\mu_k b)
\delta(\int\mu_k \beta^-)(\int\mu_k G^-)
\mu_{3g-3}(w)b(w) I^{-1}(w)|^2$$
\eqn\instop{ \prod_{l=1}^g I (v_l)Z^-(v_l) \tilde c(x_0)|^2>}
evaluated at $M_i=0$.

Since there are no available $\eta^+$'s or $\xi^+$'s to absorb
$\xi^-$'s and $\eta^-$'s,
$I(v_i)Z^-(v_i)$ can be replaced by $e^{\phi^-}
\tilde G^+(v_i)$.
In the above expression, it is easy to check that the path integral over
the $(b,c)$
ghosts cancels the path integral over the $(\beta^-,\gamma^+)$ ghosts.
The path integral over the $(\beta^+,\gamma^-)$ fields is
$$\{Z([v_1 +...+ v_g - w
 - \Delta])\}^{-1}= \{Z_1 det \omega^j (v_k)\}^{-1}$$
where $(Z_1)^{-\half}$ is the partition
function for a chiral boson and $\Delta$ is the Riemann class. Since
the $(\tilde b,\tilde c)$ path integral contributes
$$\prod_{i=1}^g \int_{a_i} Z([y_1 +...+y_g - x_0 -\Delta])=
Z_1,$$
the product of these path integrals cancels the $det \omega^j(v_k)$ factor
in \instop .

So the scattering amplitude is:
$$A_g^{2g-2}=\prod_{i=1}^g\int d^2 v_i \prod_{j=1}^{3g-3}\int d^2 m_j$$
$$
< |\tilde G^+(v_1)
...\tilde G^+(v_g)
\prod_{k=1}^{3g-4}
(\int\mu_k G^-)
\mu_{3g-3}(w)J^{--}(w)|^2>$$
which is just the topological prescription defined in \deffp .

It is easy to check that this proof for the partition function
can be repeated for $N$-point scattering amplitudes. For vertex operators
$\hat V_i=|c \exp(-\phi^+ -\phi^-)|^2 V_i$ where $V_i$ is an
N=2 primary field, the
N=2 string correlation function at instanton-number $n_I$ is
$$A_{g,1,...,N}^{n_I}
=\prod_{i=1}^g\int  d^2 M_i \prod_{j=1}^{3g-3+N}\int  d^2 m_j$$
$$
<| (\int_{a_i} \tilde b)
(\int \mu_j b))|^2\cdot$$
$$|(Z^-)^{2g-2+N+n_I}
(Z^+)^{2g-2+N-n_I}I^{n_I}\tilde c(x_0)|^2 \hat V_1(z_1)...\hat V_N(z_N)>.$$
At instanton-number $n_I=2g-2+N$, $A_{g,1,...,N}^{2g-2+N}$
is equal to the topological
expression
$$F_{g,N}^{2g-2+N}
=\prod_{i=1}^g\int d^2 v_i \prod_{j=1}^{3g-3+N}\int d^2 m_j
< |\tilde G^+(v_1)...\tilde G^+(v_g)|^2\cdot$$
$$|\prod_{k=1}^{3g-4+N}
(\int\mu_k G^-)
\mu_{3g-3+N}(w)J^{--}(w)|^2
\phi_1 (z_1) ...\phi_{N} (z_N)>$$
where $\phi_i$ is the N=2 chiral field defined by
$\phi_i=|\tilde G^+|^2 V_i$.

Although the above equivalence proof for the $N=2$ strings
and $N=4$ topological prescription has been stated at the level of amplitudes,
it is straightforward to check that the equivalence proof also holds at
the level of integrands. In other words,
$$
\prod_{i=1}^g \prod_{j=1}^{3g-4}
<| (\int_{a_i} \tilde b)
(\int \mu_j b))(\mu(w)b(w))
(Z^-)^{4g-4} I^{2g-2}(y_0)\tilde c(x_0)|^2 > $$
$$
=\prod_{i=1}^g \prod_{j=1}^{3g-4}
|\det (\omega_k(v_i))|^{-2}
 <| \tilde G^+(v_1)...\tilde G^+(v_{g})
(\int\mu_j G^-)
(\mu_{3g-3}(w) J^{--}(w))|^2 > $$
where the U(1) moduli in the non-topological expression is equal
to the divisor $[v_1 + ... + v_g - w+(2-2g)y_0+\Delta]$
in the second expression.
Since the topological integrand for unitary theories has been shown
in \unin\ to
be independent of the $v_i$ locations
for $g>1$  (or more precisely, it depends on $v_i$
like ${\rm det }(\omega_i(v_j))$),
this immediately implies for a unitary N=2 string that
the integrand
at instanton number $n_I=2g-2$ is independent of the U(1) moduli when
$g>1$.

We next consider other instanton numbers.
 Since there are total-derivative ambiguities when
both $Z^+$'s and $Z^-$'s are present in the non-topological
calculation (there is a singularity when $Z^+$ and $Z^-$
collide, so it is clear that the integrand of the non-topological
amplitude can not be independent of their locations), the integrand of
the non-topological expression can not be equal to the integrand of a
topological amplitude. However, as was shown in the previous section,
the topological partition function can be
expressed as
$$F_g^{top}=\sum_{n=2-2g}^{2g-2} {(4g-4)!\over(2g-2+n)!(2g-2-n)!}
F_g^n(u_1)^{2g-2+n} (u_2)^{2g-2-n},$$
where equation \defp\ is for $F_g^{2g-2}$. We will now prove that
the coefficients, $F_g^n$, of this polynomial provide an
unambiguous definition of the $N=2$ string partition function
on a surface of
instanton number $n$ for $2-2g\leq n\leq 2g-2$.

The first step in the proof is to start with
an N=2 string which has been rotated by the SU(2)
flavor generators, so the matter part of its
two superconformal generators are $\widehat{G^+}$ and $\widehat{G^-}$
where $\widehat{G^+}=
u^*_2 G^+ - u^*_1 \tilde G^+$ and $\widehat{G^-}=u_1 G^- + u_2 \tilde G^-$
(note that the OPE of $\widehat{G^+}$ with $\widehat{G^-}$ is the same as the
OPE
of $G^+$ with $G^-$).
Since this replaces $G^-$ by $\widehat{G^-}$ and $\tilde G^+$
by $\widehat{{\tilde G}^+}=u_1 \tilde G^+ -u_2 G^+$, it is clear
that the resulting partition function at instanton number $n_I=2g-2$ is
precisely
$F_g(u_1,u_2)$. We will now show that this partition function of the
flavor-rotated N=2 string at $n_I=2g-2$ is related to the partition
function of the unrotated string at $n_I<2g-2$.

The rotated partition function at $n_I=2g-2$ is given by the
non-topological expression:
$$\widehat A_g^{2g-2}(u_1,u_2)
=\prod_{i=1}^g\int  d^2 M_i \prod_{j=1}^{3g-3}\int  d^2 m_j$$
$$
<| (\int_{a_i} \tilde b)
(\int \mu_j b))
(\widehat Z^-)^{4g-4}
I^{2g-2}\tilde c(x_0)|^2>$$
where
$\widehat Z^-= \{\widehat Q,\xi^-\}$
and $\widehat Q$ is the flavor-rotated $Q$ which has the matter part of the
superconformal generators, $G^+$ and $G^-$, replaced with
$\widehat{G^+}$ and $\widehat{G^-}$. The next step in the proof is
to perform the unitary transformation:
$$\xi^- \to u_1 \xi^- + u_2 \xi^+ \tilde I^{-1},\quad
\eta^- \to u_1 \eta^- + u_2 \eta^+ \tilde I^{-1},$$
\eqn\gtran{\xi^+ \to u^*_2 \xi^+ - u^*_1 \xi^- \tilde I,\quad
\eta^+ \to u^*_2 \eta^+ - u^*_1 \eta^- \tilde I,}
where $\tilde I = \exp(\phi^- - \phi^+ - J)= I - c \tilde b I$ (note
that $\tilde I$ is not BRST invariant since we have removed the
$\tilde b$ dependence).
The infinitesimal version of this transformation is generated by
$$J^+ = \int \xi^+ \eta^+ \tilde I^{-1},\quad
J^- = \int \xi^- \eta^- \tilde I,$$
$$ J^3 = \int (\xi^- \eta^+ -\xi^+\eta^- +\dzm(\phi^+ - \phi^-)+
J_{color}).$$
It is easy to check that under this transformation,
$\widehat Q \to Q$, and therefore,
$$\widehat Z^- \to\{Q~,~ u_1 \xi^- + u_2 \xi^+ \tilde I^{-1}\}$$
$$= u_1 Z^- + u_2 Z^+ I^{-1} + u_2 c \xi^+ \dzm (I^{-1})$$
where we have ignored terms proportional to $\tilde b$ since they can not
contribute to the scattering amplitude.
Furthermore, since moving
the location of the $I$'s is the same as shifting the
U(1) moduli, the term proportional to $\dzm(I^{-1})$ can be dropped since
it is a total derivative in the U(1) moduli.

So after performing this unitary transformation, the rotated partition
function at $n_I=2g-2$ can be expressed as
$$\widehat A_g^{2g-2}(u_1,u_2)=
\prod_{i=1}^g\int  d^2 M_i \prod_{j=1}^{3g-3}\int  d^2 m_j$$
$$
<| (\int_{a_i} \tilde b)
(\int \mu_j b))
(u_1 Z^- + u_2 Z^+ I^{-1})^{4g-4}
I^{2g-2}\tilde c(x_0)|^2>,$$
which is equal to
$$\sum_{n=2-2g}^{2g-2} {(4g-4)!\over(2g-2+n)!(2g-2-n)!}
A_g^n (u_1)^{2g-2+n} (u_2)^{2g-2-n}$$
where $A_g^n$ is the non-rotated partition function at instanton number
$n$ which is defined in \given. We have therefore proven our
claim that the various coefficents in
the topological partition function $F_g$ provide an unambiguous
definition of the $N=2$ string partition function at all instanton
numbers. It is straightforward to generalize this proof for
$N$-point scattering amplitudes and find a similar
relation between
the topological prescription
and the $N=2$ string prescription at different instanton numbers.

There is one subtle point in this equivalence proof which needs to
be discussed.
In defining $N=2$ string scattering amplitudes, we need to integrate
over the negative-energy ghost fields $\phi^+$ and $\phi^-$. As was
shown in \ref\nega{U. Carow-Watamura, Z. Ezawa, K. Harada, A.
Tezuka and S. Watamura, Phys. Lett. B227 (1989) 73.}
 (although this reference only discusses
the N=1
ghosts, it is trivial to generalize their prescription to the N=2 case),
a convenient way to perform this integration is to sew
$\int \eta^+$ and $\int \eta^-$ around each of the $a$-cycles, and to
restrict the momentum $\int (\dzm\phi^\pm - \eta^\pm\xi^\mp$)
to be zero through the internal loops.
In order for our proof to be correct, we must check that this
prescription is unchanged by the transformation of \gtran.
Fortunately, this is the case since
$$\int
\eta^- \int \eta^+ \to
 \int (u_1 \eta^- + u_2 \eta^+ \tilde I^{-1})\int (
u^*_2 \eta^+ - u^*_1 \eta^- \tilde I)
=\int
\eta^- \int \eta^+$$
(we have used that $\int \eta^+ \int \eta^+ =0$ and that
$\tilde I$'s can be moved
around without changing the amplitude). Similarly, it is easy to check
that
$\int (\dzm\phi^\pm - \eta^\pm\xi^\mp)$ is unchanged by the transformation.

Note that the integrand of the topological amplitude
is equal
to the integrand of the $N=2$ string amplitude
 only for the terms corresponding to
the instanton numbers $n_I=2g-2$ and $n_I=2-2g$ (the $n_I=2-2g$ term
is related to the $n_I=2g-2$ term by replacing $Z^-$ with $Z^+$
and twisting the N=2 algebra in the opposite direction). For all other
$n_I$, only the integrated amplitudes are guaranteed to be
equal (this must be true since if $|n_I|<2g-2$,
the integrand of $A_g^{n_I}$ in the non-topological prescription
is only defined up to total derivatives). So even though $F_g$ is
independent of the $v_i$ locations for all values of
$u$, we are only able to say for
instanton-numbers $n_I=\pm(2g-2)$ that the non-topological
partition functions are independent of the U(1) moduli.

Since we have found a relation between flavor-rotated amplitudes
at instanton number $n_I=2g-2$, $\widehat A_g^{2g-2}$,
 and unrotated amplitudes at other
instanton-numbers, $A_g^n$, we can ask what is the relation between the
flavor-rotated amplitudes at different instanton numbers.
Let us
define $\widehat A_g^n(u,u^*)$
to be the flavor-rotated amplitude at instanton-number
$n$ (i.e., $\widehat A_g^n$ is the non-topological amplitude at $n_I=n$
using
$\widehat{G^-}$ and $\widehat{G^+}$ as the matter part of the
N=2 superconformal generators).

We can use our knowledge
of $\widehat A_g^{2g-2}$ to check that at $u_1=1$ and $u_2=0$,
\eqn\know{(u^*_a {d\over du_a})^m\widehat A_g^{2g-2} ={(4g-4)!\over (4g-4-m)!}
 A_g^{2g-2-m}}
Note that at this value of $u$, $\widehat G^+ =G^+$ and $\widehat G^-=G^-$
so $\widehat A_g^{2g-2-m}$ is equal to $A_g^{2g-2-m}$.
However, $u^*_a {d\over du_a}$ commutes with the SU(2) flavor rotations
of \flavorrot\ which transform $u_a$ and $u^*_a$ as independent doublets.
So if \know\ is valid at $u_1=1$ and $u_2=0$, it must be valid for all
values of $u$, and therefore
\eqn\also{(u^*_a {d\over du_a})^m\widehat A_g^{2g-2}
={(4g-4)!\over (4g-4-m)!}
 \widehat A_g^{2g-2-m}.}
This immediately implies that
$$
(u^*_a {d\over du_a})\widehat A_g^n =(2g-2+n)\widehat A_g^{n-1},\quad
(u_a {d\over du^*_a})\widehat A_g^n =(2g-2-n)\widehat A_g^{n+1}.$$
It is also easy to show
that the harmonicity equations defined in \harmo\ are
equivalent to the equations:
$$u^*_a {d\over dt_a} \widehat A_g^n =
u_a {d\over dt_a} \widehat A_g^{n-1},\quad
u^*_a {d\over dt^*_a} \widehat A_g^n =
u_a {d\over dt^*_a} \widehat A_g^{n-1}$$
for $2-2g<n\leq 2g-2$.

\newsec{The Self-Dual String}

In this section, we shall apply the topological method to the calculation of
scattering amplitudes for the N=2 string which describes self-dual
gravity in two complex dimensions \ov .  It will be seen that the
topological description of the amplitudes is much more convenient
to work with as there are no $N=2$ ghosts around.
  This will in particular allow us to show that all amplitudes
except for three point amplitudes vanish
to all loops.  Moreover, in the context of the self-dual string,
we will be able to interpret the $u$ variables as describing
the twistor space.

The simplest background for the $N=2$ string is $R^4$ with signature
$(2,2)$ \ref\adem{M. Ademollo, L. Brink, A. D'Adda, R. D'Auria,
E. Napolitano, S. Sciuto, E. DelGiudice, P. DiVecchia,
S. Ferrara, F. Gliozzi, R. Musto, R. Pettorino and J.H. Schwarz,
Nucl. Phys. B111 (1976) 77}\ref\dada{ A. D'Adda and F. Lizzi, Phys. Lett. B191
(1987) 85.}.
To couple a conformal theory to the $N=2$ string, we need
an $N=2$ superconformal theory, which in
particular requires a complex structure in the target-space.
  This is easily done by
viewing
$R^4$ as $C^2$, and labeling the coordinates as $(x_j,{\bar x}_{\bar
j})$ with $j=1,2$. The `Lorentz group' for this $N=2$ string is thus
the $U(1,1)$ subgroup of $SO(2,2)$ which preseves the complex structure.
The fermionic right-moving
variables of this $N=2$ string will be denoted by
$\psi^+_j$ and $\psi^-_{\bar j}$, and the fermionic left-moving variables
by $\bar \psi^+_j$ and $\bar\psi^-_{\bar j}$.
The non-zero components
of the spacetime metric are $\eta^{1{\bar 1}}=-\eta^{2{\bar 2}}=1$.

The right-moving twisted N=2 superconformal generators are given by
$$L=\dzm x_j \dzm \xbj +\pbj\dzm\pj,\quad
G^+ =\pj\dzm\xbj,\quad
G^- =\pbj\dzm\xj,\quad
J=\pj\pbj.$$
Note that after twisting,
$\pj$ has dimension 0 while $\pbj$ has dimension 1.

Since this N=2 stress-tensor has ${\hat c}=2$, it can be used to
construct an N=4 stress-tensor in the manner described in section 2.
The additional generators are
$$\tilde G^+ =\ep \pj\dzm\xk,\quad
\tilde G^- =\epb \pbj\dzm\xbk,\quad
J^{++}=\ep\pj\pk,\quad
J^{--}=\epb\pbj\pbk.$$
Note that the currents form an SU(1,1) ``color'' symmetry rather than an SU(2)
(the SU(1,1) OPE's are those of \doubalg\ and \acrdoub\ but with $J^{--}$
and $\tilde G^-$ replaced with $-J^{--}$ and $-\tilde G^-$).
Similarly, the ``flavor'' symmetry of these N=4 generators is
SU(1,1) instead of SU(2).

In defining the $N=2$ generators, we had to choose
a complex structure on $R^4$, and as is well known from twistor
constructions of self-dual metrics, the possibility for doing this is
parametrized by $SU(2)/U(1)$ for the $(4,0)$ signature and
$SU(1,1)/U(1)$ for the $(2,2)$ signature. This is precisely
the freedom we have in defining an $N=2$ algebra starting
from the $R^4$ theory which has $N=4$ symmetry (taking
into account the $\theta$ angle which was discussed in a footnote
in the previous
section).
In order to keep track of this $SU(1,1)_f$ flavor symmetry, it is
convenient to define
a 3-parameter family of N=4 generators as in section 2. These generators
are:
$$\widehat{{\tilde G}^+}(u)=u_1 \ep\pj\dzm\xk + u_2 \pj\dzm\xbj,\quad
\widehat{G^-}(u)=u_1 \pbj\dzm\xj + u_2 \epb\pbj\dzm\xbk,$$
$$\widehat{ {\tilde G}^-}(u)=u^*_2 \epb\pbj\dzm\xbk + u^*_1 \pbj\dzm\xj,\quad
\widehat{G^+}(u)=u^*_2 \pj\dzm\xbj + u^*_1 \ep\pj\dzm\xk,$$
where $|u_1|^2 - |u_2|^2 =1$. Note that we are using the notation that
$\overline{u_a} = \epsilon^{ab} \eta_{a \bar a} u^*_b$ so
$\overline{u_1}=u^*_2$ and $\overline{u_2}=u^*_1$.

Although it is not obvious in this notation, the $u$'s
keep track of not only the SU(1,1) flavor symmetry, but also the
choice of complex structure in the original
SO(2,2) spacetime.\foot{N=4 worldsheet supersymmetry was also used
by Siegel in \ref\sieg{W. Siegel, Phys. Rev. Lett. 69 (1992) 1493.}\
to describe the self-dual string in
manifest $SO(2,2)$ notation. However unlike \sieg\ , we are not attempting
to quantize the $N=4$ strings, but instead are using the $u$ variables to
choose a complex structure in the target-space, thereby allowing
quantization as an $N=2$ string. Although Siegel has claimed to find
target-space supersymmetry using his method of quantization\ref\siegtwo
{W. Siegel, Phys. Rev. D47 (1993) 2504.}, we
have yet to find such target-space
supersymmetry in the $N=2$
self-dual string.}
In order
to make this more transparent, it is convenient to write the spacetime
$x$ variables in SO(2,2) vector notation. This can be done by
defining
$x_j =x^\mu \sigma_\mu^{1 k}\epsilon_{kj}$ where
$$\sigma_0^{jk}=i\delta^{jk},\quad
\sigma_1^{jk}=i\sigma_x^{jk},\quad
\sigma_2^{jk}=i\sigma_y^{jk},\quad
\sigma_3^{jk}=\sigma_z^{jk}.$$
Note that the $\sigma^\mu$ matrices are defined such that
$\overline{(x_j)}=\xbj= x^\mu(\sigma_\mu^{2 k})\eta_{\bar j k}$.

We can therefore write the generators in SO(2,2)
notation as
$$\widehat{{\tilde G}^+}(u)=u_i\pj \dzm x^{ij},\quad
\widehat{G^-}(u)=u_i \pbj\dzm x^{ik} \epsilon_{jk},$$
$$\widehat{\tilde {G}^-}(u)=u^*_i \pbj\dzm  x^{ik}\epsilon_{jk} ,\quad
\widehat {G^+}(u)=u^*_i \pj\dzm x^{ij},$$
where $x^{ij}=x^\mu \sigma_\mu^{ij}$. Note that SO(2,2) is equivalent to
$SU(1,1)\times SU(1,1)$, and the first and second indices of $x_{ij}$
transform as doublets under the first and second SU(1,1)'s.
Therefore, just as the $u$'s keep track of how the SU(1,1) flavor
symmetry of the N=4 string is broken,
they also keep track of how the first SU(1,1) of
SO(2,2) is broken.

Although one might attempt to integrate over the $u$ variables in the
style of \harmr\ and thereby try to recover SO(2,2) invariance, this does
not seem like the right thing to do from the string point of view. Since
the $u$ variables label different choices of complex structure for the
background, integrating over the $u$'s would
be like integrating over different choices of
backgrounds for the string theory. This would be like trying to do
second-quantized string field theory, instead of following
the usual first-quantized
procedure of calculating scattering amplitudes in a fixed background.
However as discussed in the previous section, if we know
the amplitudes for a given choice
of complex structure, we can obtain the amplitudes for other choices
by the fact that the instanton corrections lead to a partition function
which transforms according to a representation of SU(2) (or (SU(1,1)),
where instanton number labels the $J_3$ eigenvalue of the flavor SU(2)
(or SU(1,1)).

Even though we have developed the above discussion in the
context of flat space, the story can be stated more generally:
Suppose we have a self-dual Ricci-flat metric on a manifold.  Then there is a
sphere (or disc in the (2,2) signature) worth of ways
to choose a complex structure on the manifold for which the Ricci-flat
metric is a K\"ahler metric, thus giving rise to an $N=2$ string
vacuum (see \ref\asmor{P. Aspinwall and D. Morrison, {\it
String Theory on $K3$ Surfaces}, preprint
DUK-TH-94-68 and IASSNS-HEP-94/23, hep-th/9404151. }\
for a recent discussion on this and its relation to $K3$ moduli).

Let us now return to the flat case and consider computing
the scattering amplitudes.
The vertex operators for the self-dual string will be constructed out of the
dimension zero primary fields
$V(k)=e^{i k^{\mu} x_{\mu}}$ where $k_\mu k^\mu=0$, and out of the
zero-momentum dimension $1/2$ chiral fields $\phi^+_j=\pj$.
As was described in the previous sections, the
topological prescription for the scattering amplitudes of these
vertex operators is given by:
$$F_{g,1,...,N}=\prod_{i=1}^g\int d^2 v_i \prod_{j=1}^{3g-3+N}\int d^2 m_j$$
$$
< |\widehat{{\tilde G}^+}(v_i)
\prod_{k=1}^{3g-4+N}
(\int\mu_k\widehat{G^-})
\int \mu_{3g-3+N} J^{--}|^2
\phi^+_1 (z_1) ...\phi^+_{N} (z_N)>$$
where for states with non-zero momentum, $\phi^+ =[\widehat{{\tilde
G}^+},V(k)]=
u_i\pj k^{ij} e^{ik^\mu x_\mu}$.

Note that for genus 0, we can use the definition of amplitudes
given in \trnpt .  In particular, the three point function is given by
$$\langle \phi^+_1 \phi^+_2 V_3 \rangle$$
which easily reproduces what one expects for the three point
function of self-dual strings \ov.
\subsec{Vanishing Theorems}
We will now discuss why certain amplitudes vanish
for the self-dual string in flat space.
The first type of vanishing theorem
will concern scattering amplitudes with
less than three external states.
The proofs will use the fact that
scattering amplitudes are invariant
under all
SO(2,2) Lorentz-transformations if the $u_i$ variables are transformed like
SU(2) spinors.
The open-string
partition function is a polynomial of degree
$4g-4$ in $u_i$ and since there are no momentum factors to contract
with the SU(2) index of $u_i$, this amplitude must vanish.\foot{
For the closed-string partition function, one has both right-handed
$u^R_i$'s and left-handed $u^L_i$'s, so it is possible to construct
the SO(2,2) scalar $(u^R_i u^L_j \epsilon^{ij})^{4g-4}$. For this reason,
we can only prove that the closed-string partition function vanishes when
$u^R_i u^L_j \epsilon^{ij}=0$ (for example, it vanishes if the
right and left-handed instanton numbers are not equal). We would like to
thank Hirosi Ooguri for pointing this out to us.}
The one-point function vanishes since the vertex operator is proportional
to the momentum, and therefore is zero by momentum conservation. For the
two-point function, there is one independent momentum $k^\mu$ satisfying
$k^\mu k_\mu=0$. Since there is no way to construct an SO(2,2) scalar out
of $k^\mu$'s, $u^R_i$'s and $u^L_i$'s which is at least linear in
$k^\mu$, the two-point function must also vanish. Note that
the vanishing theorems for
the one and two-point functions are valid for both open and
closed N=2 strings.

The second type of vanishing theorem that will be proven is that
for $N$-point functions of arbitrary genus, the amplitude vanishes unless
$k_\mu^r k^{\mu\,s}=0$ for all $1\leq r,s \leq N$
where $k^r_\mu$ is the
momentum of the $r^{th}$ external state. For three-point functions, this
is not a restriction since it is implied by
conservation of momentum and the mass-shell
condition. However for more than three external states, it implies
that the amplitude vanishes unless all particles are moving in the
same self-dual plane.  Since scattering amplitudes should be
analytic in momenta, this means that all amplitudes other than
three-point amplitudes are identically zero to all orders.

The trick to proving this second type of vanishing theorem is to
realize that the primary field $ V(k_\mu)=
e^{i k_{\mu} x^{\mu}}$ satisfies the identity
$[\int G^+, V]=h(k_\mu) [\int \tilde G^+, V]$ where $|h(k_\mu)|^2=1$.
This identity is true since
$$[\int G^+, V]=
\pj \bar k_{\bar j} e^{ik_\mu x^\mu}=h(k_\mu)
\epsilon^{jl}\pj k_l e^{ik_\mu x^\mu}
=h(k_\mu) [\int \tilde G^+, V]$$
where $h(k_\mu)=\bar k_1 /k_2$ (for convenience, we have
reverted to the original SU(1,1) notation). Note that since
$k^\mu k_\mu = k_1 \bar k_1 - k_2 \bar k_2=0$, ${\bar k_2 /k_1}=
{\bar k_1 /k_2}$.  In fact for $k\not= 0$, the $G^+$ and $\tilde G^+$
cohomology is trivial and so we can apply the construction of section
2 (see \const ) to obtain a whole ladder of observables $\phi^{(n)}$. What we
are finding here is that all $\phi^{(n)}$ constructed in this
way are proportional:
\eqn\dtn{\phi^{(n)}=h^n \phi^{(0)}}

We will now show that unless $h_r=h_s$ for $1\leq r,s\leq N$, the
amplitude vanishes. Since $h_r=h_s$ implies $\bar k^r_2/k^r_1=
\bar k^s_1/ k^s_2$ and $\bar k^r_1/k^r_2=\bar k^s_2/k^s_1$, it also
implies that $k^r_\mu k^{s\,\mu}=\bar k^r_{\bar j} k^s_j+
k^r_j \bar k^s_{\bar j}=0$.  In other words, the amplitude vanishes
unless all the
Mandelstam variables are zero. Since
for more than three external states, we can define an
amplitude with vanishing
Mandelstam variables as a limit of one with non-vanishing Mandelstam
variables, we conclude that all $N$-point functions
for $N\not=3$ vanish to all loops.  For the three-point function,
the Mandelstam variables are identically zero
so we cannot use this argument to prove vanishing, and in fact
it is known to be non-vanishing at tree \ov\
and one loop \ref\ita{M. Bonini, E. Gava and R. Iengo,
Mod. Phys. Lett. A6 (1991) 795.} .

However our symmetry
argument teaches us something also about three-point amplitudes.
We can view the momentum of a massless particle in $U(1,1)$
notation as two complex numbers of equal norm, one denoting the `energy'
and the other denoting the `momentum'.  Note moreover that the
$h$ we have defined above is simply the sum of the phases
of these two complex numbers.
For a three-point amplitude,
conservation of energy-momentum implies that we have two
equal triangles, one on the energy plane and the other on the momentum plane.
It was found in \ov\ (and similarly in \ita )
that the three point amplitude is non-vanishing
only when the two triangles have opposite orientation.  This actually
follows from the fact that $h_r$ is the same for the three particles.

We will now prove that
the amplitude vanishes when any two of the $h$'s are not equal.
Suppose that $h_1$ and $h_2$ are not equal. Then the scattering amplitude is
given by
$$F_{g,1,...,N}=\prod_{i=1}^g\int d^2 v_i \prod_{j=1}^{3g-3+N}\int d^2 m_j$$
$$
< |\prod_{i=1}^g\widehat{\tilde {G}^+}(v_i)
\prod_{k=1}^{3g-4+N}
(\int\mu_k\widehat{G^-})
(\int\mu_{3g-3+N} J^{--})|^2
\phi^+_1 (z_1) ...\phi^+_{N} (z_N)>$$
where
$$\phi^+_1 =\widehat{\tilde{\bar G}^+}\widehat{{\tilde G}^+} V_1=
\widehat{{\bar G}^+}(u_1 \tilde G^+ +u_2 G^+) e^{ik^1_\mu x^\mu}
=(u_1 h_1^{-1} +u_2)\widehat{{\tilde {\bar G}}^+} G^+ V_1 $$
and $\phi^+_2 =\widehat{{\tilde{\bar G}}^+} \widehat {{\tilde G}^+}
V_2$.

Now pull
$\widehat{{\tilde G}^+}$ off of $V_2$ and onto $J^{--}$, which is the
only place it has a singularity. The amplitude is then
$$
F_{g,1,...,N}=\prod_{i=1}^g\int d^2 v_i \prod_{j=1}^{3g-3+N}\int d^2 m_j$$
$$
(u_1 h_1^{-1} + u_2)<|\prod_{i=1}^g\widehat{{\tilde G}^+}(v_i)
\prod_{k=1}^{3g-4+N}
(\int\mu_k\widehat{G^-}) |^2
(\int\mu_{3g-3+N} \widehat {G^-}
\int\bar\mu_{3g-3+N} \bar J^{--})$$
$$
\widehat{{\tilde {\bar G}}^+} G^+ V_1(z_1)\widehat{{\tilde {\bar G}}^+}
V_2(z_2)
\phi^+_3(z_3)...\phi^+_{N} (z_N)>.$$
The next step is to pull $G^+$ off of $V_1$ and onto $V_2$. Although
$G^+$ also has singularities at the $\widehat{G^-}$'s, the residues of these
poles
are total derivatives in the $3g-3+N$ modular parameters.\foot{
In this proof of the
second type of vanishing theorem, we will ignore all potential surface
term contributions. This is presumably justified since
we can analytically continue in momenta and we have derivatives
on moduli parameters for both left- and right-movers \dixon .
The story may be more subtle for hetertotic $N=2$ strings.}

Finally, pull $\widehat {{\tilde G}^+}$ off of the last beltrami differential
and
onto $V_1$, which is the only place it has a singularity. The resulting
expression for the scattering amplitude is
$$F_{g,1,...,N}=\prod_{i=1}^g\int d^2 v_i \prod_{j=1}^{3g-3+N}\int d^2 m_j$$
$$
(u_1 h_1^{-1}+ u_2)<|\prod_{i=1}^g\widehat{{\tilde G}^+}(v_i)
\prod_{k=1}^{3g-4+N}
(\int\mu_k\widehat{G^-})
(\int\mu_{3g-3+N} J^{--})|^2$$
$$ \widehat{{\tilde {\bar G}}^+} \widehat{{\tilde G}^+}V_1(z_1)
\widehat{{\tilde {\bar G}}^+}  G^+ V_2(z_2) \phi^+_3(z_3)...\phi^+_{N}
(z_N)>.$$
Since $G^+ V_2= (u_1 h_2^{-1} +u_2)^{-1} \widehat{{\tilde G}^+} V_2$,
we get
$$F_{g,1,...,N}=
{u_1 h_1^{-1} +u_2\over u_1 h_2^{-1}+u_2} F_{g,1,..,N}.$$
By replacing $G^+$ in the above argument with $\tilde G^+$, one
can similarly prove that
$$F_{g,1,...,N}={u_1 +u_2 h_1\over u_1 +u_2h_2}F_{g,1,...,N}.$$
It is obvious that for both of these equations to be true, either
$h_1=h_2$ or $F_{g,1,...,N}$ must vanish.

Note that the above techniques can also be used to prove that
$F_{g,1,...,N}$ for the self-dual
string obeys the identity
$$ (h\, d/du_1 - d/du_2)F_{g,1,...,N}(u_1 ,u_2 ) =0,$$
and therefore $F^n_{g,1,...,N}=h F^{n+1}_{g,1,...,N}$
where $h=\bar k^r_1/k^r_2$.

\subsec {Meaning of Vanishing Theorems for Self-dual Backgrounds}

Having proven some vanishing theorems, we will now discuss
 their physical meaning\foot{Some of the above
vanishing theorems were anticipated in \dada .}.  The vanishing of an amplitude
usually implies the existence of symmetries.
It is important in this context to
recall the Coleman-Mandula theorem \ref\cm{S. Coleman and J. Mandula,
Phys. Rev. 159 (1967) 1251.}\
which states that if we have a higher-spin
symmetry in a theory, it is either a two dimensional thoery
or it is a free theory.  The idea of the proof is to look at
two-particle scattering and to show that the higher-spin
symmetry forces the amplitude to be zero except for forward
or backward scattering.  In more than two dimensions,
if the theory is not free, this would
imply a non-analytic $S$-matrix which is ruled out.
We can thus have a non-trivial theory only in two dimensions.
Indeed as is well known,
all integrable models are two-dimensional
theories of this type which
have an infinite number of higher-spin symmetries.

Here, for the $N=2$ self-dual
string, we have proven that
scattering amplitudes vanish for four and higher-point functions.
In fact, our proof resembled
the steps in the Coleman-Mandula theorem since we first showed that
the amplitudes vanish unless
the Mandelstam variables are all zero, and we then used analyticity
of the S-matrix to prove that the amplitudes always vanish.
But the self-dual $N=2$ string is nevertheless not a free theory!
The explanation for this is that
we have signature $(2,2)$ and not $(3,1)$.
For a theory with signature $(3,1)$, vanishing of the four-point amplitude
implies vanishing of the three-point amplitude
by unitarity arguments.  However for signature
$(2,2)$, we have seen that four-point amplitudes can
vanish in accord with having higher-spin symmetries
and the Coleman-Mandula theorem,
but this does not imply vanishing of the three-point amplitude (for which
there is a phase space, unlike the signature $(3,1)$ case).

We should thus expect a higher-spin
symmetry in the self-dual string. In fact, an infinite
symmetry group is expected as in two-dimensional
integrable models since most (if
not all) integrable
models in two dimensions are reductions of self-dual equations
in four dimensions.
As was discussed
in \ov , the result of \ref\pleb{J.F. Plebanski, J. Math. Phys. 16
(1975) 2395.}\ shows that
the loop group of
area-preserving diffeomorphisms is a group which maps
one solution to other solutions.  This transformation
is space-time dependent and may be viewed as corresponding to higher-spin
transformations.  It would be very important to flesh out this
symmetry in a more conventional form.  The above vanishing proof
may give some hints as to how one may realize these
symmetries more concretely.

\subsec{One loop Partition Function}

So far, we have talked about a very simple self-dual string background,
namely $R^4$.  There are other interesting classes of such backgrounds \ov ,
for example $T^*\Sigma$ (the cotangent of a Riemann surface) with signature
$(2,2)$ or Euclidean backgrounds of signature (4,0).  As for compact
Euclidean theories, there are only two possibilities: $T^4$ and $K3$.
As will be discussed in section 6, even the Euclidean backgrounds
will be important for string theory (they can be viewed as the internal part
of superstrings compactified to six dimensions).
In the Euclidean theories, the only physical fields are the moduli
of the manifold (there are no propagating modes), so the most relevant
computation is the dependence of the partition function on
the moduli.
To get a feeling for the Euclidean theories, we will consider
in this section the one loop partition function
for the case of $T^4$ (and the subcases $T^2\times T^2$,
$R^2\times T^2$ and $R^4$).  We will check at least for these cases
if it is valid to ignore the total derivative terms in deriving
\harmo\ .

For the one-loop partition function on
$T^4$, we have four left-moving and four
right-moving fermionic zero modes.
Therefore the $F_L^2$ and
$F_R^2$ in the definition of the one loop amplitude \eff\
will absorb them.  Since the fermionic and bosonic oscillators cancel
out, all that is left to contribute to \eff\ is the extrema
of classical solutions which gives the usual Narain sum \ref\nara{
K.S. Narain, Phys. Lett. B169 (1986) 41.}.
Let $\Gamma^{4,4}$ denote a self-dual $(4,4)$ Narain lattice
with a decomposition to left and right momenta $(p_L,p_R)$.
Then we have
$$F_1^{T^4}=\int d^2\tau \sum_{(p_L,p_R)\in \Gamma^{4,4}}q^{{1\over 2}p_L^2}
{\bar q}^{{1\over 2}p_R^2}$$
where as in \eff\ ,we subtract the ground state contribution
to get a convergent answer, i.e. the $(p_L,p_R)=0$
part is deleted from the sum.

For the theory with $T^2 \times T^2$,
we should restrict to lattices which decompose to $\Gamma^{2,2}
\oplus \Gamma^{2,2}= \Gamma^{4,4}$.  The moduli of each
$T^2$ is determined by a pair of complex variables $(\sigma ,\rho)$
which take values in the upper half plane modulo an $SL(2,Z)$
action and an exchange of $\sigma$ with $\rho$  \ref\vedi{R. Dijkgraaf,
E. Verlinde and H. Verlinde, {\it On the Moduli Space of Conformal
Field Theories with $c>1$}, in the Proceeding of 1987 Copenhagen
Conference `Perspectives in String Theory', World Scientific, 1988.}.
We will define
$$Z(\sigma ,\rho )=\sum_{p_L,p_R} q^{{1\over 2}p_L^2}{\bar q}^{{1\over 2}
p_R^ 2}$$
where
$$(p_L,p_R)={1\over \sqrt{2{\rm Im}\sigma{\rm Im} \rho}}(n_1+m_1\sigma +\rho
(n_2+m_2 \sigma); n_1+m_1\sigma+{\bar \rho}(n_2+m_2 \sigma ))$$
for integer $n_i,m_i$.  Then for $F_1$ on $T^2 \times T^2$ we have
\eqn\twot{F_1^{T^2\times T^2}
=\int d^2\tau \ Z(\sigma_1,\rho_1)Z(\sigma_2,\rho_2)}

We can also take the limit where the torus degenerates to obtain
the theory on $T^2\times R^2$ and $R^4$ (or $R^{2,2}$) with the results
$$F_1^{T^2\times R^2}=\int {d^2\tau \over \tau_2} Z(\sigma ,\rho )
={\rm log}(\sigma_2 \rho_2|\eta(\sigma )\eta (\rho)|^2)$$
$$F_1^{R^4}=\int {d^2\tau \over \tau_2^2}={\pi\over 3}$$
(the first integral is discussed in \ref\dkap{L. Dixon,
V.S. Kaplunovsky and J. Louis, Nucl. Phys. B355 (1991) 649.}).  It is
satisfying to see that the $N=2$ string computation of the above
in \ov\ agrees with the topological answer given by \eff .

We would now like to check the {\it harmonicity} condition
that was derived in section 2 (eq. \oneloop). This also serves to check in
at least one  concrete example whether or not there are any boundary
terms
coming from total derivatives that we have ignored in our derivation
of \harmo\ .
We will consider
the case of $T^2\times T^2$, and for concreteness, we will fix
$\sigma_1$ and $\sigma_2$ to a constant (say $i$) and concentrate on
the dependence of \twot\ on $\rho_j$ where $j=1,2$
labels the tori.

To check equation \oneloop\ , we first note that the
non-vanishing components of $M_i^{\bar j}$ and $M_{\bar i}^j$
are given by
\foot{This can be easily
derived using the $tt^*$ equations of \ref\cv{S. Cecotti and
C. Vafa, Nucl. Phys. B367 (1991) 359.}.}
$$M_{\rho_1}^{\bar \rho_2}=M_{\bar \rho_1}^{\rho_2}={{\rm Im}
\rho_2 \over {\rm Im} \rho_1}
\qquad M_{\rho_2}^{\bar \rho_1}=M_{\bar \rho_2}^{\rho_1}={{\rm Im}
\rho_1 \over {\rm Im} \rho_2}$$
{}From this, we see that equation \oneloop\ for $\partial_{\rho_1} {\bar
\partial}_{\rho_2} $ is identically satisfied
and the only equation to check is
\eqn\chk{\partial_{\rho_1}\partial_{\bar \rho_1}F_1=
{({\rm Im} \rho_2)^2\over
({\rm Im}\rho_1)^2} \partial_{\rho_2}\partial_{\bar \rho_2}F_1}
Using the identity (noted in \ref\naret{I. Antoniadis, E. Gava and
K.S. Narain, Phys. Lett. B283 (1992) 209.})
\eqn\usef{\partial_\rho {\bar \partial}_\rho Z(\sigma,\rho)={{\rm
Im}\tau\over
({\rm Im}\rho)^2}
\partial_\tau {\bar \partial}_\tau ({\rm Im} \tau \ Z(\sigma
,\rho))}
we learn that the left hand side of \chk\ is given by
$$\partial_{\rho_1}\partial_{\bar \rho_1}F_1=
\int d^2\tau {\partial_{\rho_1}}{\bar \partial}_{\rho_1}
 Z(\sigma_1,\rho_1)
Z(\sigma_2 ,\rho_2)$$
$$=\int d^2\tau {{\rm Im}\tau \over ({\rm Im}\rho_1)^2}{\partial_\tau}
{\bar \partial}_\tau ({\rm Im}\tau Z(\sigma_1,\rho_1)) Z(\sigma_2,\rho_2)$$
Now assuming that integration by parts can be done (which we will justify
in a moment), the last equation becomes
$$ {1 \over
({\rm Im}\rho_1)^2}\int d^2\tau\ {\rm Im}\tau \ Z(\sigma_1,\rho_1)
{\partial_\tau} {\bar \partial}_\tau ({\rm Im}\tau Z(\sigma_2,\rho_2) )$$
$$={({\rm Im} \rho_2)^2\over
({\rm Im}\rho_1)^2} \partial_{\rho_2}\partial_{\bar
\rho_2}F_1$$
where to get the last line, we once again use \usef , this time
in the opposite direction.

We have thus proven \chk\ modulo
checking the validity of integration by parts that we used.  This
requires checking that
\eqn\intpart{\int d^2\tau\big[ \partial {\bar \partial}({\rm Im}\tau Z_1)(
{\rm Im}\tau
Z_2)-  ({\rm Im}\tau Z_1) \partial {\bar \partial} ({\rm Im}\tau Z_2)
\big]=0.}
One has to be careful with the possible boundary contributions
because even though $Z_i$ goes to 1 exponentially
fast as $\tau \rightarrow \infty$,
the prefactor ${\rm Im}\tau\rightarrow \infty$.  To show
\intpart\ is satisfied, it is convenient to use the language
of differential forms and rewrite the left hand side of
 \intpart\ as
$$\int d
\big[\bar\partial ({\rm Im}\tau Z_1)({\rm Im}\tau Z_2)
+({\rm Im }\tau Z_1)\partial ({\rm Im} \tau Z_2)\big]$$
$$=\big[\bar\partial ({\rm Im}\tau Z_1)({\rm Im}\tau Z_2)
+({\rm Im }\tau Z_1)\partial ({\rm Im} \tau Z_2)\big]\Big|_{\tau
\rightarrow \infty}$$
$$={-1\over 2i}({\rm Im}\tau -{\rm Im}\tau
+{\rm exp. \ small})|_{\tau
\rightarrow \infty}=0.$$
This proves the validity of \intpart\ and completes the proof
of \chk .  This is a strong case for the absence of anomalies in general.
Note in particular that in the $N=2$
one-loop topological amplitude \bcov\ ,even the torus example has
anomalies.

\newsec{Topological description of the superstring}

It was recently shown that both the RNS and GS versions of the
ten-dimensional
superstring can be described by a critical N=2 string. The N=2 description
of the RNS superstring is obtained by using the results of \bv\
to embed the usual N=1 RNS string in an N=2 string.
The N=2 description of the GS superstring is related
by a field-redefinition\berko\
to this N=2 description of the RNS string
(the field-redefinition
and the N=2 $\hat c=2$ stress-tensors for these two descriptions of the
superstring can be found in Appendix A).

Since the topological methods of section 2
apply to any critical N=2
string, we can now ask what is the resulting topological prescription for
the superstring.  In one sense,
the topological prescription is closely related to the original $N=1$
RNS prescription for the superstring
since the $N=2$ ghosts
which were added to the $N=1$ RNS string
to make it into an $N=2$ string are removed in the topological theory.
Nevertheless, we shall find that the topological prescription
has an advantage over the conventional $N=1$ prescription since
it does not appear to suffer from total-derivative ambiguities.
However there is a technical point to overcome before this is fully
established. The technical point arises
because one of the N=2 matter fields in the superstring
has negative-energy (this field is related to the superdiffeomorphism
ghost of the $N=1$ superstring).

We will begin
by discussing the topological prescription for the RNS version of the
superstring. As descibed in section 2, the topological amplitude is
a polynomial in two variables which can be expressed as
$$F_g=\sum_{n=2-2g}^{2g-2} {(4g-4)!\over(2g-2+n)!(2g-2-n)!}
F_g^n (u_1)^{2g-2+n} (u_2)^{2g-2-n}.$$
We will show that $F_g^{2g-2}$ is equivalent to
the usual N=1 prescription for
RNS amplitudes when the sum of the
pictures of the vertex operators is equal to $2g-2$
(similar observations have been
recently made in \ref\marc{N. Marcus, {\it
The $N=1$ Superstring as a Topological Field Theory}, preprint TAUP-2155-94,
hep-th/9405039.}). The other $F_g^n$'s give
a new prescription for calculating RNS superstring amplitudes
when the sum of the pictures of the vertex operators is equal to $n$.
Since the topological amplitude has no total-derivative ambiguities,
these new prescriptions
may be useful in resolving the
ambiguities that plague the conventional N=1 RNS description of the
superstring.

We shall first review the conventional $N=1$ rules for calculating
RNS superstring scattering amplitudes\ref\VerNSR{E. Verlinde and
H. Verlinde, Phys. Lett. B192 (1987) 95.}. Aside from having to insert
enough $N=1$ picture-changing operators to get a non-zero result, the
only subtlety in the $N=1$ rules comes from the functional integration
over the $N=1$ superdiffeomorphism ghosts, $\beta$ and $\gamma$.
Because the $\phi$ field (which comes from bosonizing $\beta$ and $\gamma$
as $\beta=\dzm\xi\exp(-\phi)$ and
$\gamma=\eta \exp(\phi)$)
has negative-energy, the naive functional integral over its
zero modes does not converge. The explanation of this divergence is that
each physical state is represented by an infinite set of
vertex operators which are
related by picture-changing operators,
and therefore
the naive BRST cohomology contains infinite copies of each physical state.

To get a finite answer, the states propagating
in internal loops should be restricted to a fixed picture. This can be
accomplished by
inserting the operator
$\prod_{i=1}^g \delta(\int_{a_i} (\dzm\phi+\xi\eta) - P_i)$
which restricts the picture of the state propagating through the $i^{th}$
$a$-cycle to be equal to $P_i$.
Furthermore, because the $\xi$
zero mode is not present in the ($\beta$,$\gamma$) ghosts,
one should insert a $\xi$ field anywhere on the surface
and sew $\eta$ fields around each $a$-cycle in order to
reproduce the zero-mode structure of the original unbosonized system.
As was shown in \nega , these rules correctly reproduce the
usual RNS scattering amplitudes for arbitrary genus.\foot{There is
one ambiguous point in reference \nega\ which should be mentioned.
In proving that the above rules reproduce the
usual RNS scattering amplitudes, the authors of \nega\
use the fact that $\sum_{k\epsilon {\cal Z}^g}\exp(ik y)=\delta(y)$.
This is of course true if $y$ takes values on the $g$-torus
${\cal R}^g/{\cal Z}^g$, however
in this context, $y$ takes values in the Jacobian variety
${\cal C}^g/({\cal Z}^g+\tau {\cal Z}^g)$. The problem is caused by
trying to treat the right and left-moving sectors independently, which
means that one is trying to construct
a $\delta$-function which depends on $y$ but not on $\bar y$.}
It was also proven
that the resulting expressions for the scattering
amplitudes are independent of the choice of the $P_i$'s.

To get the $N=4$ topological theory for RNS strings, we define the
$N=4$ generators as
\eqn\Nfour{J^{++}=c\eta ,\quad J^{--}=b\xi ,\quad J=cb+\eta \xi,}
$$G^-=b, \quad {\tilde G^+}=\eta,$$
$$G^+=
\gamma  G_m+
c( T_m -{3\over 2}
\beta\dzm\gamma-{\half\gamma\dzm\beta}-b\dzm c )-\gamma^2 b +\partial^2 c
+\dzm (c\xi\eta),$$
$${\tilde G^-}=b (e^{\phi}G_m+\eta e^{2\phi}\partial b-c
\partial\xi)
 -\xi(T_m-{3\over 2}\beta\partial\gamma-\half\gamma
\partial\beta + 2b\partial c-c\partial b)+\partial^2 \xi$$
where $T_m$ and $G_m$ are the matter parts of the $N=1$ stress-tensor
for the RNS string with central charge $c=15$. These N=4 generators can be
obtained in the usual way from the critical
N=2 stress-tensor used to embed the RNS string in an
N=2 string. Note that $G^+=J_{BRST}+
\partial^2 c
+\dzm (c\xi\eta)$ where $Q_{N=1}=\int J_{BRST}$,
and $\tilde G^- = [Q_{N=1}, b\xi]$.

We will now show that the topological prescription for
calculating $F_g^{2g-2}$
precisely reproduces the conventional
N=1 rules for RNS scattering amplitudes when
the sum of the pictures of the vertex operators is equal to $2g-2$. Since
the N=2 matter sector of the topological
string also contains the negative-energy field $\phi$,
one gets a naive divergence from its functional integral even in the
topological prescription. This is because the cohomology of $\tilde G^+$
is trivial and as discussed in section 2,
there is therefore an infinite set of
chiral N=2 fields for each
physical state of the RNS string (note that $[G^+,W]=
[\tilde G^+,W]=0$ implies that $[G^+,W']=[\tilde G^+,W']=0$ where
$W'= \{G^+,\xi W\}$). So as in the
$N=1$ prescription, we need to restrict the picture of the
states propagating through the internal loops by inserting the operator
$\prod_{i=1}^g \delta(\int_{a_i} (\dzm\phi+\xi\eta) - P_i)$.

To calculate the term $F_g^{2g-2}$ in $F_g$,
we need to sew $G^-$'s with $3g-4$ of the Beltrami differentials,
sew $J^{--}$ with the last Beltrami differential, and
insert $[\int d^2 z (\tilde G^+
\bar{\tilde G}^+)]^g$.  To get the right-moving part of this
insertion, we should write
$$[\int d^2 z (\tilde G^+
\bar{\tilde G}^+)]^g=[\sum_{i=1}^g
(\int_{a_i}dz\tilde G^+
\int_{b_i}d\bar z\bar{\tilde G}^+
-\int_{b_i}dz\tilde G^+
\int_{a_i}d\bar z\bar{\tilde G}^+)]^g$$
$$=
g!\prod_{i=1}^g
(\int_{a_i}dz\tilde G^+
\int_{b_i}d\bar z\bar{\tilde G}^+
-\int_{b_i}dz\tilde G^+
\int_{a_i}d\bar z\bar{\tilde G}^+).$$
It is easy to check that if
$\tilde G^+$ is holomorphic and $\bar{\tilde G}^+$ is anti-holomorphic,
this last expression is proportional to
$det (Im\tau)|\prod_{i=1}^g \int_{a_i} dz \tilde G^+|^2$, and therefore
the holomorphic part is obtained by sewing $\tilde G^+$ around the $g$
$a$-cycles.\foot{The $det(Im\tau)$ factor is cancelled by the
functional
integral over the $\phi$ and $(\xi,\eta)$ fields which contributes
$det(Im\tau)^{-1}$. Because the restriction on the $\phi$ momentum
removes the quadratic part of the soliton action for these fields, there
is no compensating contribution of $det(Im\tau)^{+1}$ coming
from the sum over soliton configurations. This explains why the functional
integral contains dependence on $Im\tau$, unlike the functional integrals for
ordinary chiral bosons.}
But using the N=4 generators defined in \Nfour\ ,
$\tilde G^+=
\eta$, $G^-=b$, and $J^{--}=b\xi$. We therefore
reproduce the N=1 prescription of \nega\ where the $\xi$ field
is inserted on one of the Beltrami differentials. Since there are no
explicit N=1
picture-changing operators in the topological prescription, all $\phi$
charge must come from the vertex operators which means that the sum of their
pictures must equal $2g-2$.

The condition on the sum of the pictures of the vertex operators
can be understood by observing that $G^\pm$ carries zero picture while
$\tilde G^\pm$ carries $\pm 1$ picture (the picture-counting operator is
$R= \int (\dzm\phi+\xi\eta)$). Therefore for $F_g^n$ to be non-zero,
$$\sum_{r=1}^N p_r  =(n-g+2)-1 +(g-1)=n$$
where $p_r$ is the picture of the $r^{th}$ vertex operator,
$(n-g+2)$ is the number of $\tilde G^+$'s minus the number of $\tilde G^-$'s
in the calculation of $F_g^n$, $-1$ is the picture of $J^{--}$, and $(g-1)$
is the anomaly in $R$ on a genus $g$ surface. We therefore
conjecture that $F_g^n$ provides an unambiguous definition of the
RNS scattering amplitude when the sum of the pictures of the vertex
operators is equal to $n$.

One point which needs to be addressed is the functional integral over the
negative-energy $\phi$ field. For $F_g^{2g-2}$, this functional integral was
regularized by inserting around the internal loops the operator
\eqn\opins{\delta(\int (\dzm\phi+\xi\eta) - P_i)=\delta(R-P_i)}
for some integer or half-integer $P_i$. For the general $F_g^n$,
it is not obvious that
this regularization will give the correct scattering amplitude.
Although $R$ commutes with $G^\pm$, it does not commute with
$\widehat{ G^\pm}$
and is therefore not invariant under SU(2) flavor rotations.

However it will now be argued that the precise form of $R$ is irrelevant
for the calculation, and different choices for $R$ give the same
result. The only requirement on $R$ is that the $\delta$-function insertion
fixes the momentum of the
$\phi$ field, thereby reducing the cohomology
to a single copy of each physical state. The argument is that since
the scattering amplitude has been proven to
be independent of the choice of the $P_i$'s
 we can replace $P_i$ by the momentum of any field
other that $\xi\eta$ or $\phi$, and the amplitude will not change.
For example, we can replace $P_i$ by $\int bc+P_i'$, so the
new regularization is given by inserting the operator
$$\delta(\int (\dzm\phi+\xi\eta-bc)-P_i').$$
If we do the functional integration over the $\phi$ and $(\eta,\xi)$
fields before we do the integration over the $(b,c)$ fields, it is clear
that we can treat the zero mode $\int bc$ as a constant.

In fact, we could now reverse the roles of $(b,c)$ and $(\eta,\xi)$
and first do the functional integration over the $\phi$ and $(b,c)$ fields.
Since $\int \eta\xi$ now acts as a constant, we can remove this constant
by shifting $P_i'\to P_i'-\int\xi\eta$. So the resulting regularization
prescription would be to insert the operator
$$\delta(\int (\dzm\phi-bc)-P_i')=\delta(R' - P_i').$$
Note that $R'=\int(\dzm\phi -bc)$ satisfies the commutation relations
$[R',G^\pm]=\pm G^\pm$ and $[R',\tilde G^\pm]=0$, so it is related to
$R$ by an SU(2) flavor rotation ($R'$ is the $N=1$ ghost-number operator).
So we have argued that the precise form of the regularization for the
$\phi$ field does
not affect the scattering amplitude, and we can therefore
use the operator insertion of \opins\ for all $F_g^n$.
To be honest, we should note that the above argument is only as rigorous
as the ambiguous point of reference \nega\ which was
mentioned in a previous footnote. It would therefore be nice to
perform explicit calculations using different regularizations
for the $\phi$ field and to explicitly check if the
resulting expressions are independent of the choice of regularization.

We will now discuss the topological prescription for the GS version of the
superstring. Since we will also discuss the GS superstring
in the following section, we will limit our discussion here to
those aspects that are not covered later in the paper.
For further details on the
non-topological description of the uncompactified
ten-dimensional GS superstring, see
\ber ,while for details on the GS superstring
compactified on a Calabi-Yau manifold, see reference \ref\bbbb{N. Berkovits,
{\it Covariant Quantization of the Green-Schwarz Superstring in a
Calabi-Yau Background},
preprint KCL-TH-94-5, hep-th/9404162.}.

Since the RNS description of the superstring is related to
the GS description by a field redefinition, all aspects of the
topological prescription are easily translated from RNS language
into GS language.
For example, as shown in Appendix A,
the picture-counting operator
in the RNS description is translated into the $R$-parity
operator in the GS description. This means that $F_g^n$ in the GS topological
prescription calculates the term in the scattering amplitude which has
an $R$-anomaly equal to $n$.
Therefore after compactifying down to four dimensions,
the scattering amplitude of $N^+$ chiral fermions, $N^-$ anti-chiral fermions,
and $M$ vector bosons is given by $F_g^n$ where $n=\half(N^+ -N^-)$
(note that the $g^{th}$-order term in this scattering
amplitude vanishes if $|N^+ -N^-|>g-1$).
For compactifications down
to six dimensions, $R$-parity is
defined by breaking the internal SU(2) of the spinors down to U(1).
As will be discussed in the following section, this preserves the
SO(5,1) Lorentz invariance but breaks half of
the eight spacetime supersymmetries. For the uncompactified superstring
in ten dimensions, $R$-parity is defined by breaking the SO(9,1)
Lorentz invariance down to $SU(4)\times U(1)$.

The only element of the GS topological prescription that requires further
discussion is the functional integration over the negative-energy field.
In the four and six-dimensional compactifications, this field was called
$\rho$, while in the uncompactified GS superstring, it was called
$(h^+ -h^-)$. As in the RNS topological prescription, we need to regularize
the functional integral by inserting a delta-function of
an operator that restricts the momentum of the negative-energy field
through the internal loops. However, we should not use a delta-function
of the
$R$-parity operator
(which is related to the RNS picture-counting operator
by the field-redefinition)
since this operator does not
commute with spacetime supersymmetry. Instead, we should choose an
operator which commutes with spacetime supersymmetry
but still restricts the momentum of the negative-energy field.\foot
{In previous work on non-topological
multiloop calculations for the uncompactified
GS superstring, it was shown that the functional integral over the
$h^\pm$ fields was well-defined if the U(1) moduli, $M_i$, was restricted to
satisfy the condition that $M_i =[\sum_{r=1}^m c_r z_r]_i$
where
$z_r$ are the locations of the fields $\exp(c_r h^-(z_r))$ and
$[x-y]_i=\int_x^y \omega_i$. This condition on the U(1) moduli can
be understood as coming from the insertion of the operator
$\prod_{i=1}^g \delta(\int_{a_i} \dzm h^+ - M_i)$
which restricts the momentum of
the negative-energy field through the internal loops. Note that because
the negative-energy field in the GS formalism transforms under
U(1) transformations,
its momentum is not integer or half-integer
valued, but is shifted from an integer value by the U(1) moduli.}
For the
reasons stated earlier, the resulting
scattering amplitude should not depend on our choice,
however it would be nice to verify this fact
with an explicit comparison using different operators.

\newsec{Topological amplitudes and superstring compactifications}

It was previously shown that for the superstring compactified on a
six-dimensional Calabi-Yau manifold, certain spacetime scattering amplitudes
corresponding to superpotential terms
can be computed exactly and can be expressed as topological amplitudes
of the $(N=2,\hat c=3)$ superconformal field theory representing
the Calabi-Yau manifold \bcov \nar . Since we now have a prescription for
calculating
topological amplitudes of an $(N=4,\hat c=2)$ superconformal field theory, it
is natural to conjecture that these new topological amplitudes correspond to
certain spacetime scattering amplitudes
coming from superpotential terms when the superstring is
compactified on the four-dimensional $K3$ manifold (or $T^4$
for the one loop correction). As will be
shown in this section, the conjecture is correct.

Because the target-space described by the superstring on $K3$ is
spacetime supersymmetric in six dimensions, it is convenient
to use six-dimensional Green-Schwarz variables to perform the computation
of the superstring scattering amplitudes. These Green-Schwarz variables
are related to the conventional Ramond-Neveu-Schwarz variables by a
field redefinition which can be found in Appendix A. Since this field
redefinition preserves all OPE's, the
computations using the Green-Schwarz variables should be equivalent to the
computations using the Ramond-Neveu-Schwarz variables. The
GS description of the superstring contains N=2 worldsheet
supersymmetry, so one might ask how it can be equivalent to the N=1
RNS description of the superstring. The answer is that the N=1 RNS
superstring (or any critical N=1 string) can be embedded in
an N=2 string\bv , and the GS description is related by the above
field redefinition to the
N=2 version of the RNS string \berko
(in fact, this was how the $N=1\to N=2$ embedding was discovered).

There are three advantages to using the GS variables instead of the
RNS variables. Firstly, because
the GS variables are all GSO projected (i.e., they have no square-root
cuts with the spacetime-supersymmetry generators), there is no need to
sum over spin structures when performing multiloop computations.
Secondly, all states which are independent of the compactification
moduli (i.e.,
the graviton, gravitino, etc.) can be
represented by vertex operators constructed entirely out of spacetime
GS fields. And thirdly, manifest spacetime-supersymmetry can be preserved
at all stages in the calculation.

Because the GS techniques will not be familiar to most readers, we
will begin by reproducing the Calabi-Yau topological amplitudes
as a warm-up exercise for the $K3$ case. In other words, we will show
using four-dimensional GS techniques
that a $g^{th}$ order term in the
scattering amplitude of $2g$
four-dimensional
supermultiplets is given by a topological amplitude of the $(N=2,
\hat c=3)$
superconformal field theory representing the Calabi-Yau manifold.
After completing this warm-up exercise, we will show using
six-dimensional GS techniques that a $g^{th}$ order term in the scattering
amplitude of $4g$ six-dimensional
supermultiplets is given by a topological amplitude
of the $(N=4,\hat c=2)$ superconformal field theory representing the $K3$
manifold.
We will only discuss the Type II closed string case in which the
supermultiplets describe supergravity fields, however it should be
possible to generalize to the heterotic case where the supermultiplets
describe super-Yang-Mills fields.

\subsec{GS superstring on CY 3-fold}
As was discussed in \bbbb ,
the worldsheet variables of the four-dimensional GS superstring
consist of the spacetime variables, $x^m$ ($m=0$ to 3), the right-moving
fermionic
variables,
$\t^\a$ and $\tb^\ad$ ($\a,\ad=1$ to 2), the conjugate right-moving
fermionic variables, $p_\a$ and $p^*_\ad$, and one right-moving
boson $\rho$.
The chiral boson
$\rho$ is identified with $\rho \pm 2\pi$ (in two-dimensional
Minkowski space, $\rho$ is imaginary valued and $i\rho$ takes values
on a circle of radius 1) and is related
to R-transformations of four-dimensional superspace.
For the Type II GS superstring, one also has
the left-moving fermionic fields,
$\th^\a$,$\th^{*\ad}$,
$\bar p_\a$, $\bar{ p^*}_\ad$, and one left-moving
boson, $\bar\rho$.
In this section, an asterisk will mean anti-chiral whereas
a bar will mean left-moving.

In conformal gauge, the worldsheet action for these fields is:
$$\int d^2 z [\half\dzp x^m \dzm x_m + p_\a \dzp\t^\a +
p^*_\ad \dzp\tb^\ad +\half\dzp \rho\dzm\rho +
\bar p_\a \dzm\th^\a +
{\bar p^*_\ad} \dzm\th^{*\ad} +\half\dzp \bar\rho\dzm\bar\rho ].$$
The free-field OPE's for these worldsheet variables are
$$x^m(y) x^n(z)\to -\eta^{mn}\log|y-z|,
\quad \rho(y) \rho(z) \to -\log(y -z),$$
$$p_\a(y)\theta^\b (z)\to {\delta_\a^\b\over{y -z}},\quad
 p^*_\ad(y)\tb^\bd (z)\to {\delta_\ad^\bd\over{y -z}}, $$
$$ \bar p_\a(y)\bar\theta^\b (z)\to {\delta_\a^\b\over{\bar y -\bar z}},\quad
{\bar p}^*_\ad(y){\bar\tb}^\bd (z)\to
{\delta_\ad^\bd\over{\bar y -\bar z}},\quad
\bar\rho(y) \bar\rho(z) \to -\log(\bar y -\bar z).$$
Note that the chiral boson $\rho$ can not
be fermionized since
$e^{\rho(y)}~e^{\rho(z)}~\to e^{2\rho(z)}(y -z)^{-1}$ while
$e^{\rho(y)}~e^{-\rho(z)}~\to (y -z)$. It has the same behavior as the
negative-energy field $\phi$ that appears when bosonizing the RNS ghosts
$\gamma=\eta e^{\phi}$ and $\beta=\partial\xi e^{-\phi}$\ref\FMS{D.
Friedan, E. Martinec and S. Shenker, Nucl. Phys. B271 (1986) 93.}.

These worldsheet GS variables form a representation of an $N=2$
superconformal algebra with $c=-3$. The generators of this algebra are
given by:
\eqn\GSf{L_{d=4}=\half\dzm x^m \dzm x_m +
p_\a\dzm \t^\a +  p^*_\ad \dzm\tb^\ad +\half\dzm\rho\dzm\rho}
$$G^-_{d=4}=e^{\rho} (d)^2 , \quad
G^+_{d=4}=e^{-\rho} ( d^*)^2, \quad
J_{d=4}=\dzm\rho, $$
where
$$d_\a=p_\a+i\tba\dzm x_{\a\ad}-\half(\tb)^2\dzm\t_\a
+{1\over 4}\t_\a \dzm (\tb)^2,$$
$$ d^*_\ad= p^*_\ad
+i\ta\dzm x_{\a\ad}-\half(\t)^2\dzm\tb_\ad
+{1\over 4}\tb_\ad \dzm (\t)^2, $$
and $(d)^2$ means
$\epsilon^{\a\b} d_\a d_\b$. It is straghtforward to check
\ref\siegGS{W. Siegel, Nucl. Phys. B263 (1986) 93.}\
that $d_\a$ and $d^*_\ad$ commute with the
$(N=2,d=4)$ spacetime supersymmetries which are generated by
\eqn\GSsusy{
q_\a=\int dz [p_\a -i\tba\dzm x_{\a\ad}-{1\over 4}(\tb)^2\dzm\t_\a],}
$$ q^*_\ad=\int dz^- [ p^*_\ad
-i\ta\dzm x_{\a\ad}-{1\over 4}(\t)^2\dzm\tb_\ad],$$
$$\bar q_\a=\int d\bar z [\bar
p_\a -i\tba\dzp x_{\a\ad}-{1\over 4}(\th^*)^2\dzp\th_\a],$$
$$
{\bar q}^*_\ad=\int d\bar z [{\bar p}^*_\ad
-i\th^\a\dzp x_{\a\ad}-{1\over 4}(\th)^2\dzp\th^*_\ad].$$
The generator of $R$-parity transformations in superspace is given by:
\eqn\Rparity{R=\int dz (\dzm\rho+\half(p_\a \t^\a -p^*_\ad \t^{*\ad})),}
$$\bar R=\int d\bar z (\dzp\bar\rho+\half(\bar p_\a \bar\t^\a -
\bar p^*_\ad \bar\t^{*\ad})).$$ As will be shown in
Appendix A, the $R$-weight of a GS vertex operator
is equal to the picture of the corresponding RNS vertex operator.
Therefore the sum of the $R$-weights of the GS vertex operators is
equal to the instanton-number of the surface that contributes to the
scattering amplitude.
It is easy to check that the N=2 tensor of equation \GSf\
commutes with the spacetime-supersymmetry and $R$-parity generators.

Since the Calabi-Yau manifold is descibed by an N=2 superconformal
field theory with $c=9$,
the combined system
of the four-dimensional GS superstring and the Calabi-Yau manifold
is described by an N=2 superconformal field theory with $c=6$. The
generators of the corresponding N=2 algebra are given by:
\eqn\GSfour{L_{GS}= L_{d=4}+~L_{CY},\quad
G^-_{GS}=G_{d=4}^- + ~G^-_{CY}, }
$$G^+_{GS}=G_{d=4}^+~+ G^+_{CY}, \quad
J_{GS}=J_{d=4}~+J_{CY},$$
where $[L_{CY},G^-_{CY},G^+_{CY},J_{CY}]$ are the $(N=2,c=9)$
generators describing the Calabi-Yau manifold and
$[L_{d=4},G^-_{d=4},G^+_{d=4},J_{d=4}]$ are the $(N=2,c=-3)$
generators defined in \GSf\ .

As was described in reference \bbbb , this N=2 tensor is
related by a field redefinition to the N=2 tensor obtained by embedding
the N=1 RNS string into an N=2 string (see appendix A). An advantage
of the GS variables is that it allows the N=2 tensor to be split into
two pieces, one of which is independent of the internal manifold.
As will be shown below, this simplifies the computation of scattering
amplitudes.

All physical states of the superstring are represented
by vertex operators of the form
$\hat V=|c  e^{-\phi^+-\phi^-}|^2 V $
where $V$ is an N=2 primary field which is constructed entirely out
of matter fields and is dimension (0,0). In other words, $L$ and $G^\pm$
have only $(y -z)^{-1}$ singularities with $V$,
$\bar L$ and $\bar G^\pm$
have only $(\bar y -\bar z)^{-1}$ singularities with $V$,
while $J$ and $\bar J$ have no singularities with
$V$. To obtain vertex operators in other pictures,
one can attach arbitrary combinations of $Z^\pm$, $\bar Z^\pm$, $I^\pm$,
and $\bar I^\pm$ onto $\hat V$.

For example, if $V$ depends only on the four-dimensional GS fields and is
independent of the Calabi-Yau manifold,
$$|Z^- Z^+|^2 \hat V= |c~ ( d^{*\ad} ~\N^2\Nba  +
d^a~\Nb^2\Na +$$
$$\dzm\tba~ \Nba +
\Pi^{\a\ad}~\Na\Nba )~+ \gamma~ e^{-\rho}\bar d^\ad ~\Nba |^2 V
$$
where $\Na V= [\int dz~ d_\a~,V] $
and
$\Nba V= [\int dz ~ d^*_\ad~,V]$.
Since $c\bar c$ can be replaced with $\int d^2 z $, the vertex operator
can be written in integrated form as:
\eqn\integrated{U=
\int d^2 z ~
|\bar d^\ad ~\N^2\Nba  +d^a ~\Nb^2\Na}
$$ +\half(
\dzm\tba~ \Nba +\dzm\ta~\Na + \Pi^{\a\ad}~(\Na\Nba-\Nba\Na) )|^2 V.
$$

For the
massless supermultiplet of the closed superstring which describes
the four-dimensional supergravity fields, the vertex operator is
constructed from the primary field
$V(\t,\t^*,\bar\t,\bar\t^*,x) $. The constraints of being N=2 primary
imply that
$$(\N)^2 V=(\N^*)^2 V=(\bar\N)^2 V=(\bar\N^*)^2 V=\partial^m\partial_m V=0$$
where $\Na=\partial_{\t^\a} + i\t^{*\ad} \partial_{\a\ad}$,
$\nabla^*_\ad=\partial_{\t^{*\ad}} + i\t^{\a} \partial_{\a\ad}$,
$\bar\nabla_\a=\partial_{\bar\t^\a} + i\bar\t^{*\ad} \partial_{\a\ad}$,
$\bar\nabla^*_\ad=\partial_{\bar\t^{*\ad}} + i\bar\t^{\a} \partial_{\a\ad}$.
The gauge transformations of $V$ which leave the integrated vertex operator
unchanged are
$$\delta V=
(\N)^2\Lambda +(\N^*)^2\Lambda^*
+(\bar\N)^2\bar\Lambda +(\bar\N^*)^2 \bar\Lambda^*.$$
The remaining unconstrained components describe the physical massless
fields of the superstring which are independent of the Calabi-Yau
manifold. For example, the
graviton and axion are described by $g_{mn}+b_{mn}
=\sigma_m^{\a\ad}\sigma_n^{\b\bd}
\Na\N^*_\ad\bar\N_\b\bar\N^*_\bd V$, the chiral graviphoton
by $T_{\a\b}=(\N^*)^2\Na (\bar\N^*)^2 \bar\N_\b V$, etc.).

Since the GS formalism is a critical N=2 string theory,
we can use \corfu\ to
calculate superstring scattering amplitudes. Therefore the genus $g$
scattering amplitude for $2g$ massless supermultiplets is given by:
\eqn\ampfour{A_g= |\int du \sum_{n_I=2-2g}^{2g-2}
(u^*_2)^{2g-2+n_I} (u^*_1)^{2g-2-n_I}|^2}
$$ \prod_{j=1}^{3g-3}\int d^2 m_j
\prod_{i=1}^g \int d^2 v_i
<|\prod_{i=1}^{g-1} \widehat{{\tilde G}^+}(v_i) J(v_g)
(\int\mu_j \widehat{G^-})|^2
U_1 ... U_{2g}>,$$
where
$$U_i=\int  d^2 z
|d^{*\ad} ~(\N)^2\Nba +d^a ~(\Nb)^2\Na $$
$$ +\half(
\dzm\tba~ \Nba +\dzm\ta~\Na + \Pi^{\a\ad}~(\Na\Nba-\Nba\Na) )|^2 V_i,$$
\eqn\Gdef{
\widehat{{\tilde G}^+} = u_1 (e^{2\rho+J_{CY}} (d)^2 +e^{\rho}
\tilde G^{++}_{CY})
+
u_2 ( e^{-\rho} ( d^*)^2+G^+_{CY}),}
$$\widehat{G^- }= u_1(e^{\rho}(d)^2 + G^-_{CY}) +
u_2 ( e^{-2\rho-J_{CY}} ( d^*)^2+e^{-\rho}\tilde G^{--}_{CY}),$$
$\tilde G^{\pm\pm}_{CY}$ is the residue of the pole in the OPE of
$e^{\pm J_{CY}}$ with $G^\mp_{CY}$, and $U$ comes from
\integrated\ . The integration over $u$
with the insertion of
$(u^*_2)^{2g-2+n_I} (u^*_1)^{2g-2-n_I}$
picks out the contribution of instanton number $n_I$ to the amplitude since
up to an overall normalization,
\eqn\factor{\int du  (u_1^{2g-2+m} u_2^{2g-2-m})
(u^*_2)^{2g-2+n} (u^*_1)^{2g-2-n} =
\delta_{mn}(2g-2+m)!
(2g-2-m)!.}
So the combinatorial factor in \homp\ is appropriately cancelled.
Summing over $n_I$ includes the contributions of all possible
$R$-parity anomalies to the scattering amplitude with the instanton
$\theta$-angle set equal to zero (it is clear that shifting the
$\theta$-angle is equivalent to rotating the vertex operators by a phase
proportional to their $R$-weight).
Note that the integration over $u$ is done independently in the
right and left-moving
sectors.

The claim is that the piece of this amplitude that describes the scattering
of $2g-2$ chiral graviphotons and two gravitons is
\eqn\claim{
\int d^2  \t d^2\bar\t (W_{\a\b} W^{\a\b})^{g} T_{CY}}
where $W_{\a\b}$
is the chiral field strength of $(N=2,d=4)$ supergravity
(the lowest component of
$W_{\a\b}$ is the chiral graviphoton and the $\t\bar\t$ component
is the Riemann tensor) and
$T_{CY}$ is the topological partition function of the Calabi-Yau
manifold.
Note that it makes sense to define a chiral superspace integral in
\claim\ since $\N_\ad W_{\a\b}=\N_\ad T_{CY}=0$. The chirality
of $T_{CY}$ comes from the holomorphicity condition which says
that (up to possible anomalies), $T_{CY}$ is independent of $\bar t_i$
if the action is deformed as in \deform\ .
Since the massless
deformations of the Calabi-Yau manifold described by $t_i$
are four-dimensional chiral scalar superfields, $\N_\ad T_{CY}=
(\N_\ad t_i)\partial_{t_i} T_{CY}=0$.

The first step in proving \claim\
is to realize that since the chiral graviphoton carries $R$-weight
$+\half$ and the graviton carries $R$-weight 0, the relevant term in
\ampfour\ is the
piece with $n_I=g-1$ (recall that the sum of the
$R$-weights of the vertex operators is equal to the instanton number
of the surface that contributes to the scattering amplitude). We therefore
have $3g-3-m$ $G^-$'s, $m$ $\tilde G^-$'s, $g-1-m$ $G^+$'s, and
$m$ $\tilde G^+$'s contributing in \ampfour\ . Furthermore, because of the
background charge, these operators need to contribute
$1-g$ units of
$\rho$ charge and
$3g-3$ units of $J_{CY}$ charge. It is easy to check that this is
only possible if each $G^-$ contributes $G_{CY}^-$, each $\tilde G^-$
contributes
$e^{-2\rho-\int J_{CY}} (d^*)^2$, each $G^+$ contributes $(d^*)^2 e^{-\rho}$,
and each $\tilde G^+$ contributes $\tilde G_{CY}^{++} e^{\rho}$.

It is convenient to regularize the functional integral over
the negative-energy
by constraining $\int (\dzm\rho-
p^*_1 \theta^{*1})$ to be zero through the internal loops.
This has the effect of fixing the $v_i$'s such that the ones
that go with $\tilde G^+$ are sewn in with the $m$ beltrami
differentials that contribute $\tilde G^-$. The remaining $g-m$ $v_i$'s
can be inserted at arbitrary points on the surface (with this choice of
the $v_i$ locations, $\rho$ always occurs in the combinations
$\rho-i\sigma$ where $i\dzm\sigma=p^*_1 \t^{*1}$).
The resulting expression for the
$n_1=g-1$ piece of \ampfour\ is:
$$F_g^{1-g}=
\prod_{j=1}^{3g-3}\int d^2 m_j det(Im\tau) |det \omega^k(v_i)|^2
$$
$$<|\prod_{i=m+1}^{g-1} e^{-\rho}(d^*)^2 (v_i) J(v_g)
\prod_{j=1}^m (\int\mu_j G^-_{CY} e^{-\rho}(d^*)^2)
\prod_{k=m+1}^{3g-3}(\int\mu_k G^-_{CY})|^2$$
$$U_1 ... U_{2g}>,$$
where $v_1 ... v_m$ coincide with the first $m$ beltrami differentials.
The
\eqn\insertion{det(Im\tau)|det \omega^k(v_i)|^2}
factor comes from the fact that
the $\rho$ regularization contributes
$$det(Im\tau)\prod_{i=1}^g \delta([\sum_j a_j y_j -\sum_k b_k z_k]_i)$$
where $[y-z]_i=\int_z^y \omega_i$,
$y_j$ is the position of $e^{a_j\rho(y_j)}$ and
$z_k$ is the position of $e^{i b_k \sigma(z_k)}$ (note that
$\t^{*1}=e^{-i\sigma}$ and $p^*_1=e^{i\sigma}$).
It is easy to check that only the $\int d^2 z_i |d^a (\Nb)^2\Na|^2 V$=
$\int d^2 z_i d^\a \bar d^\b W_{\a\b}$ term contributes
in $U_i$ of \integrated\
because we need at least $2g$ zero modes
of $d^\a$ and $\bar d^\a$.

To perform the functional integral over the $\rho$, $d^{*\ad}$ and
$\t^{*\ad}$ fields, it is useful to pull $G^+$ off one of the vertex
operators and circle it around $J(v^g)$ (otherwise the naive functional
integral is ill-defined since one gets 0/0). The resulting expression for
the scattering amplitude is:
$$F_g^{1-g}=
\prod_{j=1}^{3g-3}\int d^2 m_j det(Im\tau) |det \omega^k(v_i)|^2
$$
$$<|\prod_{i=m+1}^{g} e^{-\rho}(d^*)^2 (v_i)
\prod_{j=1}^m (\int\mu_j G^-_{CY} e^{-\rho}(d^*)^2)
\prod_{k=m+1}^{3g-3}(\int\mu_k G^-_{CY})|^2$$
$$
(\prod_{r=1}^{2g-1}\int d^2 z_r d^\a \bar d^\b W_{\a\b})
\int d^2 z_{2g} |e^{\rho} d^a~\Na |^2 V>.$$

It is easy to check that
the functional integrals over the $\rho$ and
$(p^*_1, \theta^{*1})$ fields
contribute $|Z_1|^{-2}[det(Im\tau)]^{-1}$ where
$(Z_1)^{-\half}$ is the partition function for a chiral boson. The functional
integral over the
$(p^*_2,\theta^{*2})$ fields contributes $|Z_1 det \omega_j(v_k)|^2$.
The $(p_\a,\t^\a)$ functional
integrals, when integrated over the
vertex operator locations, give an overall factor of $|Z_1|^4
(det[Im\tau])^2$ and
contract the spinor indices of the $W$'s.
Finally, the functional integral over the four $x^m$'s give a factor
of $|Z_1|^{-4}(det[Im\tau])^{-2}$. Putting all of this together, one finds
$$F_g^{1-g}=
\int d^2\t d^2\t^* d^2\bar\t d^2\bar\t^*
(W^{\a\b} W_{\a\b})^{g-1} W^{\g\d}
\N_\g \bar \N_\d V$$
$$\prod_{j=1}^{3g-3}\int d^2 m_j
<| \prod_{j=1}^{3g-3} \int\mu_j G^-_{CY}|^2 >,
$$
where the superspace integration comes from the zero mode of the functional
integral over the $\t$'s and $\t^*$'s. Performing the integrations over
$\t^*$ and $\bar\t^*$ (which turns $\N_\g \bar\N_\d V$
into $W_{\g\d}$), one proves the claim that
\eqn\holomorphic{ F_g^{1-g}=
\int d^2  \t d^2\bar\t (W_{\a\b} W^{\a\b})^{g} T_{CY}}
where $T_{CY}=
\prod_{j=1}^{3g-3}\int d^2 m_j
<| \prod_{j=1}^{3g-3} \int\mu_j G^-_{CY}|^2 >.$

\subsec{GS superstring in $K3$ Background}

After completing the warm-up exercise with a Calabi-Yau
background, we are now ready to use GS techniques to
study superstring amplitudes in a $K3$ background.
In the Calabi-Yau case, we were able to
preserve manifest the SO(3,1) Poincar\'e invariance, as well as the
8 spacetime supersymmetries of the four-dimensional Type II superstring.
In the $K3$ case, we will be able to preserve manifest the full
SO(5,1) Poincar\'e invariance, but only 8 of the 16
spacetime-supersymmetries of the six-dimensional Type II superstring.
Although it may be possible to modify the six-dimensional GS formalism
so that it preserves manifest all 16 of the spacetime supersymmetries,
we do not see how to do this at the present time. Nevertheless, the
six-dimensional GS formalism is still much more convenient than the
RNS formalism for the reasons stated at the beginning of this section.

The worldsheet variables of the six-dimensional GS superstring
consist of the spacetime variables, $x^m$ ($m=0$ to 5), the right-moving
fermionic
variables, $\t^\a$ ($\a=1$ to 4), the conjugate right-moving
fermionic variables, $p_\a$, and two right-moving
bosonic variables,
$\rho$ and $\sigma$. Under the SO(5,1) super-Poincar\' e
transformations, $m$ is a vector index whereas $\a$ is a spinor
index (note that a vector index can be represented as an anti-symmetric
product of two spinor indices). As in the four-dimensional GS superstring,
$\rho$ is imaginary-valued like the $\phi$ boson in the bosonized
N=1 superconformal ghosts. The chiral boson $\sigma$, on the other hand,
is real-valued like that obtained from bosonizing a pair of chiral fermions.
The background charge in the untwisted theory is $+2$ for both of these
chiral bosons (after twisting $L\to L+\half \dzm J$, the background
charge for these bosons is $+3$).
For the Type II GS superstring, one also has
the left-moving fermionic variables,
$\th^\a$,
$\bar p_\a$, and the left-moving
bosonic variables, $\bar\rho$, $\bar\sigma$. The field redefinition that
relates these six-dimensional GS variables
to the conventional RNS variables can be found in Appendix A.

In conformal gauge, the worldsheet action for these fields is:
$$
\int d^2 z [\half\dzp x^m \dzm x_m + p_\a \dzp\t^\a +
\half\dzp \rho\dzm\rho -
\half\dzp \sigma\dzm\sigma +$$
$$
\bar p_\a \dzm\th^\a +
 +\half\dzp \bar\rho\dzm\bar\rho
-\half\dzp \bar\sigma\dzm\bar\sigma ].$$
The free-field OPE's for these worldsheet variables are
$$x^m(y) x^n(z)\to -\eta^{mn}\log|y-z|,$$
$$ \rho(y) \rho(z) \to -\log(y -z),\quad
\quad \sigma(y) \sigma(z) \to +\log(y -z),$$
$$p_\a(y)\theta^\b (z)\to {\delta_\a^\b\over{y -z}}, \quad
 \bar p_\a(y)\bar\theta^\b (z)\to {\delta_\a^\b\over{\bar y -\bar z}},$$
$$
\bar\rho(y) \bar\rho(z) \to -\log(\bar y -\bar z), \quad
\bar\sigma(y) \bar\sigma(z) \to +\log(\bar y -\bar z).$$

In terms of these GS variables, the untwisted
$c=6$ N=2 stress-tensor is:
$$L_{GS}=\half\dzm x^m \dzm x_m +
p_\a\dzm \t^\a +\half\dzm\rho\dzm\rho
-\half\dzm\sigma\dzm\sigma
+\partial_z^2 (\rho+i\sigma)+~L_{K3}$$
$$G^-_{GS}=e^{-i\sigma-2\rho} (p)^4+ ~
e^{-\rho} \epsilon_{\a\b\g\d}p^\a p^\b \dzm x^{\g\d} +
$$
$$e^{i\sigma}(
\dzm x^m \dzm x_m +
2 p_\a\dzm \t^\a  +\dzm(\rho+i\sigma)\dzm(\rho+i\sigma))
+~
G^-_{K3},$$
$$ G^+_{GS}=e^{-i\sigma} ~+ G^+_{K3},$$
\eqn\GSsix{J_{GS}=-\dzm(\rho+i\sigma)~+J_{K3},}
where
$(p)^4$ means
$\epsilon^{\a\b\g\d} p_\a p_\b p_\g p_\d$, and $[L_{K3},G^-_{K3},G^+_{K3},
J_{K3}]$ forms a $c=6$ N=2 stress-tensor which describes the
$K3$ manifold. As in the Calabi-Yau case, this $(N=2,c=6)$ stress-tensor
is related to the N=2 tensor of the RNS string by a field redefinition
(see Appendix A), and the GS version has the advantage
that it splits the N=2 tensor into a
six-dimensional piece and a $K3$ piece.

The spacetime supersymmetry generators can be written as:
$$q^1_\a=\int dz ~p_\a ,
\quad  q^2_\a=\int dz [e^{-i\sigma-\rho} p_\a
-i\e\t^\b\dzm x^{\g\d}].
$$
$$\bar q^1_\a=\int d\bar z ~\bar p_\a ,
\quad \bar q^2_\a=\int d\bar z [e^{-i\bar\sigma-\bar\rho} \bar p_\a
-i\e\bar\t^\b\dzp x^{\g\d}].
$$
which satisfy the anti-commutation relations
$\{q^i_\a~,~q^j_\b\}=\epsilon^{ij}
\e \int dz ~\dzm x^{\g\d}$ and
$\{\bar q^i_\a~,~{\bar q}^j_\b\}=\epsilon^{ij}
\e \int d\bar z ~\dzp x^{\g\d}$.
Because there is no internal SU(2) index on $\t^\a$ and $\bar\t^\a$,
only half of the spacetime supersymmetries
are manifestly preserved in the six-dimensional
GS formalism (the spacetime supersymmetries generated by $q^1_\a$
and $\bar q^1_\a$ are manifest since $\t^\a$ and $\bar\t^\a$ only appear
in the form $\dzm\t^\a$ and $\dzp\bar\t^\a$).
Note however that scattering
amplitudes will of course
be invariant under all of the spacetime supersymmetries.
The reason why the internal SU(2) cannot be manifestly
preserved in the N=2 GS
string is because a U(1) direction has to be chosen to define
$R$-parity.
The generators of the six-dimensional $R$-parity
transformations are:
\eqn\Rparitysix{R=\int dz [\dzm \rho+\half p_\a \t^\a],}
$$\bar R=\int d\bar z [\dzp \bar\rho+\half \bar p_\a \bar\t^\a].$$
Note that the N=2 tensor \GSsix\ is invariant under all of the
spacetime supersymmetries and $R$-parity transformations.

As in the four-dimensional case, all physical states of the six-dimensional
GS superstring can be represented by vertex operators of the form
$\hat V=|c e^{-\phi^+-\phi^-}|^2 V$ where $V$ is an N=2 primary field.
For the massless fields independent of the $K3$ background,
$V$ is constructed just out of the
six-dimensional GS fields.
Because it has dimension (0,0) and has no singularities
with $J$, $V$ only depends on $\sigma$ and $\rho$ in the combinations
$$V=\sum_{m,n=-\infty}^{+\infty} e^{m(i\sigma+\rho)+n(i\bar\sigma+\bar\rho)}
V_{m,n} (x,\t,\bar\t) .$$
Furthermore, since $V$ has only a pole singularity with $G^+$ and
$\bar G^+$, $V_{m,n}$=0
for $m>1$ or $n>1$.

Vertex operators in integrated form are obtained by hitting $\hat V$ with
$Z^+ Z^-$ and removing the $c$ ghost. Since $\{\bar G^+,[G^+, V]\}
=e^{\rho+\bar\rho}V_{1,1}$,
it is easily shown that the integrated form of the above vertex operators
is:
\eqn\integratedsix{U=
\int d^2 z |e^{-i\sigma-\rho} \eu p_\a ~(\NB\Ng\Nd )+
 \eu p_\a ~(\NB \partial_{\g\d} )}
$$
+e^{i\sigma+\rho}
(\dzm x^m ~\partial_m +\dzm\ta~\Na )|^2 V_{1,1}$$
where $\Na={d\over d\ta}$.
Note that $[G^+,V]$ having only a single pole with $G^-$ implies the
on-shell constraints
\eqn\onshell{(\N)^4 V_{1,1}=\eu \Na\NB\partial_{\g\d} V_{1,1}=
\partial_{m}\partial^{m}V_{1,1}=0.}
The remaining unconstrained components of $V_{1,1}$ describe the
physical massless fields of the superstring which are independent of the
$K3$ moduli. For example, the graviton and axion are described by
$g_{mn}+b_{mn}
=\sigma_m^{\a\b}\sigma_n^{\g\d}
\Na\N_\b\bar\N_\g\bar\N_\d V_{1,1}$, and the four graviphotons
by $u^{*i} \bar u^{*j} T_{ij}^{\a\bar \a}= |\eu(u^*_1\N_\b\N_\g\N_\d
+u^*_2 \N_\b\partial_{\g\d})|^2 V_{1,1}$.
Note that the four graviphotons
have $(\pm\half,\pm\half)$ $R$-parity, so we can define the chiral
graviphoton
to be $T_{22}^{\a\bar \a}$ which has $(+\half,+\half)$ $R$-parity.

Suppose we are scattering $4g$ supermultiplets on a genus
$g$ surface. Then the $N=4$ topological amplitude is
\eqn\ampsix{A= |\int du \sum_{n_I=1}^{2g-2}
(u^*_1)^{2g-2+n_I} (u^*_2)^{2g-2-n_I}|^2}
$$\prod_{j=1}^{3g-3}\int d^2 m_j
\prod_{i=1}^g \int d^2 v_i
<|\prod_{i=1}^{g-1} \widehat{{\tilde G}^+}(v_i) J(v_g)
(\int\mu_j \widehat{G^-})|^2
U_1 ...U_{4g} >,$$
where
$$U_i=\int d^2 z_i |e^{-i\sigma-\rho} \eu p_\a ~(\NB\Ng\Nd )+
 \eu p_\a ~(\NB \partial_{\g\d} )+$$
$$
+e^{i\sigma+\rho}
(\dzm x^m ~\partial_m +\dzm\ta~\Na )|^2 V_{1,1},$$
$$\widehat{{\tilde G}^+}= u_1 e^{J_{K3}}( e^{-3\rho-2i\sigma} (p)^4+ ~
e^{-2\rho-i\sigma} \epsilon_{\a\b\g\d}p^\a p^\b \dzm x^{\g\d} +
$$
$$e^{-\rho}(
\dzm x^m \dzm x_m +
2 p_\a\dzm \t^\a  +\dzm(\rho+i\sigma)\dzm(\rho+i\sigma)))$$
$$
+~
u_1 e^{-\rho-i\sigma}\tilde G^+_{K3}+u_2 (e^{-i\sigma} ~+ G^+_{K3}),$$
$$\widehat{G^-}= u_1
( e^{-2\rho-i\sigma} (p)^4+ ~
e^{-\rho} \epsilon_{\a\b\g\d}p^\a p^\b \dzm x^{\g\d} +$$
$$
e^{+i\sigma}( \dzm x^m \dzm x_m +
2 p_\a\dzm \t^\a  +\dzm(\rho+i\sigma)\dzm(\rho+i\sigma))+
G^-_{K3})$$
\eqn\defG{+u_2 (e^{-J_{K3}+\rho} +e^{\rho+i\sigma}\tilde G^-_{K3}) .}

The claim is that the piece of this amplitude that
describes the scattering of $4g-4$ chiral graviphotons
and 4 gravitons is
\eqn\claimsix{\int d^4
\t_+ \int d^4{\bar\t_+} \int du \int d\bar u
(\epsilon\bar\epsilon)^g [\widehat W_{++}^{\a\bar\a}]^{4g}
F_g(u,\bar u)}
where $\t_-^\a=u^b \t_b^\a$ and
$\t_+^{\a}=u^{*b} \t_b^\a$ are the usual harmonic superspace
variables constructed out of the SO(5,1) spinors
with internal SU(2) index $\t_b^\a$ and $\t_b^{*\a}$ (similarly
for ${\bar \t_-^\a}$
and ${\bar \t_+^{\a}}$),
$\widehat W_{++}^{\a\bar\a}(u^*,\bar u^*,\t_+,\bar\t_+)$
is the analytic
field strength in harmonic superspace which satisfies the
on-shell conditions
\eqn\Wonshell{{d\over d\t_-^\b}\widehat W^{\a\bar\a}
={d\over d\bar\t_-^{\b}}\widehat W^{\a\bar\a}=
u^*_b {d\over du_b}\widehat W^{\a\bar\a}=
\bar u^*_b {d\over d\bar u_b}\widehat W^{\a\bar\a}=0}
(the lowest component of $W_{++}$ is $u^*_i \bar u^*_j$
times the graviphoton $T_{ij}$ and
the $\t_+{\bar\t_+}$ component is
the Riemann tensor),
the $(4g,4g)$ spinor indices of $[W]^{4g}$ are contracted with
$g$ $\e$'s and $g$ $\bar\epsilon_{\bar\a\bar\b\bar\g\bar\d}$'s in all
possible ways,
and $F_g(u,\bar u)$ is the
topological partition function of \homp\ for the $N=2$ string
on the $K3$
manifold.

It will now be shown that the harmonicity condition \harmo\ for
$F_g$ allows the harmonic superspace integral of \claimsix\ to
make sense.
First note that $F_g$ is a function of degree $(4g-4,4g-4)$ in
$(u,\bar u)$, so it carries U(1) number $(-4g+4,-4g+4)$. When
combined with the $(4g,4g)$ U(1) number of the $W$'s and the
$(-4,-4)$ U(1) number of the $\t_+ \bar\t_+$ integration, the
harmonic integrand has the correct U(1) charge.
Second, note that the superspace integral only makes sense if
\eqn\second{{d\over d\t_-^\b} \int du \int d\bar u
(\epsilon\bar\epsilon)^g [\widehat W_{++}^{\a\bar\a}]^{4g}
F_g(u,\bar u)}
is zero.
Because of \harmo\ (we are ignoring possible anomalies),
$${d\over dt^i_a}F_g=
{d\over du_a} Y^i_g$$
for some $Y^i_g$ where $t^i_a$ are the six-dimensional
scalar hypermultiplets that describe the massless deformations of the
$K3$ manifold (the equation of motion for the scalar hypermultiplet is
$\partial_{\t_a^\b} t^i_b =\delta^a_b f^i_\b$ for some $f^i_\b$).
Using the equations of motion for $W_{++}$ in \Wonshell\ ,
\second\ is equal to
$$ \int du \int d\bar u
(\epsilon\bar\epsilon)^g [\widehat W_{++}^{\a\bar\a}]^{4g}
{d\over d\t_-^\b}F_g$$
$$= \int du \int d\bar u
(\epsilon\bar\epsilon)^g [\widehat W_{++}^{\a\bar\a}]^{4g}
(u^*_a {d\over d\t_a^\b}t^i_b) {d\over dt^i_b}F_g$$
$$= \int du \int d\bar u
(\epsilon\bar\epsilon)^g [\widehat W_{++}^{\a\bar\a}]^{4g}
u^*_a f^i_\b {d\over du_a}Y^i_g$$
\eqn\third{= \int du \int d\bar u (u^*_a {d\over du_a})
(\epsilon\bar\epsilon)^g [\widehat W_{++}^{\a\bar\a}]^{4g}
f^i_\b Y^i_g.}
But by the rules of harmonic integration, \third\ is zero so
we have checked that the harmonic superspace integral in \claimsix\
makes sense.

To prove \claimsix\ , one first uses the fact that
the chiral graviphoton has $R$-weight $(+\half,+\half)$ and the
graviton has $R$-weight 0, and therefore
the only term that contributes is
that of instanton number $(n_I=2g-2,\bar n_I=2g-2)$.
This means that we have just $G^-$'s and $\tilde G^+$'s, and since
we need
$2g-2$ units of $J_{K3}$ charge,
the only
contribution from $\widehat{G^-}$ is $u_1 G_{K3}^-$ and the only
contribution from $\widehat{{\tilde G}^+}$
is $u_1 e^{-\rho-i\sigma} \tilde G^+_{K3}$.
In order that $\tilde G^+_{K3}$ can not be
pulled off the surface, $J(v_g)$ must contribute $J_{K3}$. Also,
in order to have enough $p_\a$'s and enough $e^{\rho+i\sigma}$'s,
only the term
$$\int d^2 z_i |\e p_\a (\N_\b\partial_{\g\d})|^2 V_{1,1}=
\int d^2 z_i p_\a \bar p_{\b} W^{\a\b}_{2,2}$$
contributes to $U_i$ of
\integratedsix\ .
Note that $W_{22}^{\a\b}$ is defined as $W_{++}^{\a\b}$ when
$u^*_1=0$, $u^*_2=1$, $\t_+^\a=\t^\a$, and $\bar\t_+^\a=\bar\t^\a$.

So the relevant term in the scattering amplitude is
$$F_g^{2-2g}=
\prod_{j=1}^{3g-3}\int d^2 m_j
\prod_{i=1}^g \int d^2 v_i
$$
$$<|\prod_{i=1}^{g-1} e^{-\rho-i\sigma}\tilde G^+_{K3} (v_i) J_{K3}(v_g)
\prod_{j=1}^{3g-3}(\int\mu_j G^-_{K3})|^2
\prod_{r=1}^{4g}
\int d^2 z_i p_\a \bar p_{\b} W^{\a\b}_{22}>.$$

It is convenient to regularize the functional integral over the
negative-energy $\rho$ field by constraining $\int \dzm(\rho+i
\sigma)$ to be zero through the internal loops.\foot{
Since $\rho$ only occurs in the combination $\rho+i\sigma$, the locations
of the $v_i$'s are not restricted by this regularization. So we can
either remove the integration over the $v_i$'s and insert $det(Im\tau)
|det \omega^k(v_i)|^2$ as in \insertion , or we can leave the integration
unchanged. These two prescriptions are equivalent as was shown in \unin .}
With
this regularization, the functional integrals over $\rho$ and
$\sigma$  contribute $|Z_1|^{-2}[det(Im\tau)]^{-1}$ where
$(Z_1)^{-\half}$ is the partition function for a chiral boson.
The functional integral over
the six $x$'s give a contribution of $|Z_1|^{-6}(det [Im\tau])^{-3}$, and the
functional integral over the $p^\a$'s and $\t^\a$'s (after integrating over
the locations of the vertex operators) contributes $|Z_1|^8(det [Im\tau])^4$
and a product of $g$ $\e$ $\eb$'s. Putting all this together, one finds
\eqn\onepiece{F_g^{2-2g}=
\prod_{j=1}^{3g-3}\int d^2 m_j
\prod_{i=1}^g \int d^2 v_i
\int d^4\t  d^4\bar\t}
$$<|\prod_{i=1}^{g-1} \tilde G^+_{K3} (v_i) J_{K3}(v_g)
\prod_{j=1}^{3g-3}(\int\mu_j G^-_{K3})|^2>
(\epsilon\bar\epsilon)^g [W^{\a\bar\a}_{22}]^{4g},$$
where the superspace integration comes from the zero mode of the functional
integral over the $\t$'s and $\bar\t$'s.

Since $W^{\a\bar\a}_{22}$ is just one component of
$W^{\a\bar\a}_{++}$, internal SU(2) invariance implies that
$(W^{\a\bar\a}_{22})^{4g}$ appears as one of the components of
$(W_{++}^{\a\bar\a})^{4g}$.
{}From equation \ampsix\ , it will now be argued
that the scattering amplitude of \onepiece\
is one component of the amplitude
\eqn\proofclaim{\int d^4 \t_+ \int d^4\bar\t_+ \int du \int d\bar u
(\epsilon\bar\epsilon)^g [W_{++}^{\a\bar\a}]^{4g}}
$$\prod_{j=1}^{3g-3}\int d^2 m_j
\prod_{i=1}^g \int d^2 v_i
<|\prod_{i=1}^{g-1} \widehat{{\tilde G}^+_{K3}} (v_i) J_{K3}(v_g)
\prod_{j=1}^{3g-3}(\int\mu_j\widehat{G^-_{K3}})|^2>,$$
thereby proving the claim of \claimsix\ .

The other contributions to $W_{++}^{\a\bar\a}$ come from the term
$$e^{-i\sigma-\rho}\eu p_\a (\N_\b \N_\g \N_\d)$$
in $U_i$ of
equation \integratedsix\ . In other words,
\eqn\defW{W_{++}^{\a\bar\a}=
|u^*_1 (\NB\Ng\Nd )+
 u^*_2 (\NB \partial_{\g\d} )|^2 V_{1,1}}
where $\t_+^\a$ is defined as $u^*_2 \t^\a$ (note that all $\t^\a$
dependence comes from the $u^*_2$ term because $(\N)^4 V_{1,1}=0$
by the on-shell condition of \onshell\ ).

Because non-chiral graviphotons can
have $-\half$ $R$-parity, the scattering amplitude for $4g$
$W_{++}$'s involves
surfaces of all instanton numbers. However, let us consider only the
following contributions of $\widehat{{\tilde G}^+}$ and $\widehat{G^-}$
in \defG\ :
$$\widehat{{\tilde G}^+}= u_1 \tilde G^+_{K3} e^{-\rho-i\sigma} +u_2 G^+_{K3},
\quad
\widehat{G^-}=u_1 G^-_{K3} + u_2 \tilde G^-_{K3} e^{\rho+i\sigma}$$
(when $n_I\neq |2-2g|$, the
other pieces of $\widehat{{\tilde G}^+}$ and $\widehat{G^-}$ may contribute
to terms in \ampsix\ which are unrelated to the
scattering of $4g-4$ chiral graviphotons).
Since the functional integrals over the six-dimensional
GS fields are the same as in the $W^{\a\bar\a}_{11}$ case (the
$e^{\pm(\rho+i\sigma)}$ factors cancel each other out), the
only remaining part of \proofclaim\ that needs to be explained is the
$u$ dependence.

We have to show that the harmonic integration over $u$
gives the correct scattering amplitude at each value of the
$R$-anomaly.
If the sum of the $R$-parities of the vertex operators is equal to $n_I$,
then we need a factor of $(u^*_1)^{2g-2-n_I} (u^*_2)^{2g-2+n_I}$
in order to select $F_g^n$ from the topological amplitude
(recall \factor\ ).
But since each graviphoton in $W^{\a\bar\a}$
contributes $|(u^*_1)^{\half-R}(u^*_2)^{\half+R}|^2$ where
$R$ is its $R$-weight, the integration over $u$ assigns the
correct $n_I$ to the scattering amplitude.

\newsec{Conclusion}
We have formulated a new topological string theory based
on $N=4$ superconformal symmetry
which has critical dimension $\hat c=2$.  Each such topological
theory comes in a family parametrized by $S^2=SU(2)/U(1)$ and the
partition function
at each genus $g$ forms a representation of spin $2g-2$
for this $SU(2)$.
Since the $N=2$ string with matter and ghosts decoupled is equivalent
to this new topological theory, we can use the topological
reformulation to gain insight into the structure of some $N=2$
string vacua.  There are two basic classes of examples of $N=2$
string vacua: One corresponds to four-dimensional self-dual geometries
which in signature $(2,2)$ has only one physical propagating particle.
The other class of examples corresponds to the ten-dimensional superstring
which, in both the RNS and GS formulations, has a twisted $N=4$ superconformal
symmetry with $\hat c=2$.

Using the $N=4$ topological description, we proved certain vanishing
theorems for the $N=2$ self-dual string on $R^4$.  In particular, we
showed to all loops that only the three-point function and the closed-string
partition function are non-vanishing,
which strongly hints at an integrable structure for the self-dual string.
We also studied two Euclidean self-dual
Ricci-flat backgrounds: $T^4$ and
$K3$.  It was shown that the partition
function of the $N=2$ string on $K3$ computes superpotential terms
in harmonic superspace generated by compactifying
the type II superstring to six dimensions
on $K3$.  This is very much analogous to the fact
that $N=2$ topological theories on Calabi-Yau compute
superpotential terms generated by compactifying the superstring to
four dimensions on a
Calabi-Yau manifold.

We also used the topological reformulation of superstrings
to show how the $N=4$ topological string
theory provides a definite
prescription for computation which may resolve the question of ambiguities
for RNS strings.

There are a number of things left to do:  Even though we have
formulated the computation of the $N=2$ string partition function in
a convenient
topological language, we do not yet know how to compute
it explicitly.  We believe that this should be possible analogous
to what occurs for the Kodaira-Spencer theory of gravity.
Another question to resolve is the {\it mathematical
meaning} of the
topological partition function we are computing. For the partition function
of $N=2$ topological strings,
we are counting the number of holomorphic maps
from the Riemann surface to the Calabi-Yau in one version,
while in another
version, we are studying a quantum variation of Hodge structure.
Similar interpretations would be desirable for $N=4$ topological
strings.  However the situation
is bound to be more complicated simply because it
includes both versions of the $N=2$ topological theory at the same time.

Another issue worth investigating is whether or not there are
analogs of holomorphic anomalies for $N=4$ topological strings.
We explicitly checked one example ($T^4$) where such anomalies could have
existed, but we found that they are absent.

As far as RNS strings, we have found a definite
prescription for computation which appears to avoid ambiguities.  It would
be interesting to explicitly calculate the superstring partition function
using this new prescription and
see how the $N=4$ topological reformulation removes the
ambiguities.

Another application of the new topological theory is to shed
new light on
the relationship between the worldsheet supersymmetries and target-space
supersymmetries of the GS superstring. It has been conjectured by many
physicists that the correct description of the
GS superstring in $d$ spacetime dimensions should be with a string
theory containing $d-2$ worldsheet supersymmetries\ref\manyGS
{D. Sorokin, V. Tkach, D. Volkov and A. Zheltukhin,
Phys. Lett. B216 (1989) 302\semi
N. Berkovits, Phys. Lett. B241 (1990) 497\semi
M. Tonin, Phys. Lett. B266 (1991) 312\semi
F. Delduc, A. Galperin, P. Howe and E. Sokatchev,
Phys. Rev. D47 (1993) 578\semi
L. Brink, M. Cederwall and C. Preitschopf, Phys. Lett. B311 (1993)
76.}.
In such a
description, all of the
local fermionic Siegel-symmetries (which are necessary to get to
light-cone gauge) can be substituted with worldsheet supersymmetries.
Until now, this description was realized at the quantum level only for
the $N=2$ worldsheet-supersymmetric description of the four-dimensional
GS superstring\bbbb . But with the results of this paper, the $N=4$
worldsheet-supersymmetric description of the six-dimensional GS
superstring can now also be realized at the quantum level. Of course,
the most interesting case of the ten-dimensional GS superstring
is yet to be solved, however it suggests that there may be an
$N=8$ topological string theory waiting to be discovered.

It is quite possible that our new topological theory leads to other
insights.  For example, if one considers self-dual strings,
it has been speculated \ov\ that there is a twisted supersymmetry in the
target-space which would explain why the physical modes are concentrated
on self-dual backgrounds as appears in four-dimensional topological
theories \ref\witf{E. Witten, Comm. Math. Phys. 117 (1988) 353\semi
E. Witten, {\it Supersymmetric Yang-Mills Theory on a Four Manifold},
preprint IASSNS-HEP-94/5, hep-th/9403195.}.
In the topological reformulation of the $N=2$ string, one indeed
has a twisted $N=1$ spacetime supersymmetry in the target space.\foot
{Because this spacetime supersymmetry is twisted, it does not
generate new physical fermionic states. However, the twisted supersymmetry
algebra may be realized in the string field theory by including the ghost
fields.}
The four supercharges are the zero modes of
$G^+$, ${\tilde G}^+$, and $\psi_{\bar i}$, which have
the non-vanishing anti-commutators
$$\{ G^+,\psi_{\bar i} \}=\partial x_{\bar i}$$
$$\{ \tilde G^+, \psi_{\bar i} \}=\epsilon_{\bar i \bar j}\eta^{j\bar
j}\partial x_{
j}.$$
It is amusing that the most successful approach in computations
for four-dimensional topological theories has been based on considering
K\"ahler backgrounds (which is automatic for $N=2$ strings),
and deforming the theory while preserving $N=1$ topological symmetry
(which is similar to what one has here for $N=2$ strings).
Clearly, many interesting discoveries remain to be made in this area.

{\bf Acknowledgements}: We would like to
thank Michael Bershadsky,
Paul Howe, Chris Hull, Nobuyoshi Ohta, Hirosi Ooguri, Alvaro
Restuccia, Warren Siegel,
Dmitri Sorokin, and
Paul Townsend for useful discussions, and N.B. would
like to thank Harvard University for its
hospitality, and the SERC for its financial support.
The research of C.V. was supported in part by the Packard fellowship
and NSF grants PHY-92-18167 and PHY-89-57162.

\newsec {Appendix A: The relationship between the GS and RNS formalisms}

In this appendix, we will describe the relation between the GS formalism
and the RNS formalism of the superstring. Because the GS superstring has
N=2 superconformal invariance, it is related to the N=2 embedding of the
N=1 RNS superstring. As was discussed in reference \bv , this embedding
is described by the following (N=2,c=6) stress-energy tensor
constructed out of the RNS fields:

$$L=\half\dzm x^m\dzm x_m +\psi^m \dzm \psi_m + L_{RNS} -{3\over 2}
\beta\dzm\gamma-{\half\gamma\dzm\beta}-{3\over 2}b\dzm c +\half c\dzm b
+\half \dzm(\xi\eta),$$
\eqn\RNStensor{G^-=b,}
$$G^+=\gamma (\psi^m \dzm x_m + G_{RNS}^+ +G_{RNS}^-)+
$$
$$c(\half\dzm x^m\dzm x_m +\psi^m \dzm \psi_m + L_{RNS} -{3\over 2}
\beta\dzm\gamma-{\half\gamma\dzm\beta}-b\dzm c )-\gamma^2 b +\partial_-^2 c
+\dzm (c\xi\eta),$$
$$J=cb+\eta\xi,$$
where $x^m$ and $\psi^m$ are the spacetime RNS
matter fields, $[b,c]$ and
$[\beta=\dzm \xi e^{-\phi},\gamma=\eta e^{\phi}]$
are the twisted RNS ghost fields, and $[L_{RNS},
G_{RNS}^+,$
$G_{RNS}^-, J_{RNS}]$ is the N=2
stress-energy tensor coming from the compactification (we will assume
that the compactification preserves at least $(N=1,d=4)$ spacetime
supersymmetry so the internal manifold has $N=2$ worldsheet
supersymmetry).

There is an OPE-preserving field redefinition
which maps GS fields into GSO-projected combinations of RNS fields
(GSO-projected means that there are no square-root cuts with the
spacetime-supersymmetry generators) and maps the
N=2 tensor of the GS string into the N=2 tensor of equation \RNStensor\ .
Since the N=2 tensor of equation \RNStensor\
contains both N=1 RNS matter and
ghost fields, this field redefinition maps GS matter fields into
RNS matter and ghost fields. This is expected since spacetime-supersymmetry
transforms GS matter fields into matter fields
but mixes RNS matter and ghost fields together.

Using the N=2 rules of computation (either in the non-topological
form with N=2 ghosts, or in the
topological form discussed in this paper), the two superstring
formalisms should give the same scattering amplitudes since they are
related by a field redefinition. However, the actual calculations will
look different since
the GS fields are GSO projected before doing the
path integral.\foot{A similar situation exists in the GS and
RNS light-cone
formulations of the superstring \ref\lcGS
{M. Green and J. Schwarz, Nucl. Phys. B243 (1984) 475\semi
S. Mandelstam, Prog. Theor. Phys. Suppl. 86 (1986) 163\semi
A. Restuccia and J.G. Taylor, Phys. Rep. 174 (1989) 283.}.
Although a field-redefinition
based on SO(8) triality relates the
light-cone GS fields with the light-cone RNS fields, the amplitude
calculations
look very different in the two formulations since in the GS light-cone
formulation, there are no square-root cuts and no sum over spin-structures.}
Therefore
in the GS formalism, there are no square-root cuts at any
stage in the calculation and there is no need to sum over spin structures.
This allows the calculations to be performed in a manifestly
spacetime-supersymmetric manner. Up to now in ten-dimensional GS
calculations, it has only been possible to
manifestly preserve
an $SU(4)\times U(1)$ subgroup (or an
$SO(5,1)\times U(1)$ subgroup) of the original $SO(9,1)$ Lorentz-invariance.
However, if the compactification
itself breaks the $SO(9,1)$ down to $SO(5,1)\times U(1)$ or smaller
(as in the
Calabi-Yau compactification to four dimensions or the
$K3$ compactification to six dimensions where the internal
$SU(2)$ is broken to $U(1)$), this is not a disadvantage.

We will first discuss the relationship between GS and RNS fields for
the four-dimensional case when the superstring is compactified on a
Calabi-Yau manifold, and will then discuss the relationship for the
six-dimensional case when the superstring is compactified on $K3$.
For the ten-dimensional uncompactified case, see reference \berko .

For the four-dimensional case,
first define a ``chiral'' set of GS variables by
performing the unitary transformation,
$$\tilde\Phi=\exp (-\int dz [i\dzm x_{\a\ad} \ta\tba
+ e^{-\rho}(\t)^2 G_{CY}])$$
$$
\Phi~ \exp(\int dz [i\dzm x_{\a\ad} \ta\tba
+ e^{-\rho}(\t)^2 G_{CY}]),
$$
where $\Phi$ includes all GS fields defined in Section 6.1. In terms of
these chiral GS variables, $G^+_{GS}+G^+_{CY}$ is
simply $e^{\tilde\rho} (\tilde p)^2$.
The field redefinition from these chiral GS variables to the RNS
variables is:
\eqn\fourredef{\tilde x^m_{GS}=x^m_{RNS},\quad
\dzm\tilde\rho=-3\dzm\phi+cb+2\xi\eta-J_{CY}^{RNS},}
$$ \tilde\t^\a=c \xi e^{-\half(3\phi+
\int^z J_{CY}^{RNS})} S^\a,
\quad
\tilde\t^{*\ad}=e^{\half(\phi+
\int^z J_{CY}^{RNS})}S^\ad,$$
$$\tilde{p_\a}=b \eta e^{\half(3\phi+
\int^z J_{CY}^{RNS}])}S_\a,\quad
\tilde{p^*_\ad}=e^{-\half(\phi+
\int^z J_{CY}^{RNS})}S_\ad,$$
$$\tilde L_{CY}=L_{CY}^{RNS}+{3\over 2}(\dzm\phi+\eta\xi)^2-
(\dzm\phi+\eta\xi)J_{CY}^{RNS},\quad
\tilde G_{CY}
=e^{\phi}\eta G_{CY}^{RNS}$$
$$ \tilde{\bar G}_{CY}
=e^{-\phi}\xi \bar G_{CY}^{RNS},
\quad
\tilde J_{CY}=J_{CY}^{RNS}+3(\dzm\phi+\eta\xi),$$
where $S^\a$ is the SO(3,1) chiral spinor constructed out of the
four $\psi$ fields as $S^\a=e^{\half\int^z
(\pm \psi^0\psi^1\pm\psi^2\psi^3)}$
with an even number of $+$ signs, and $S^\ad$ is the
anti-chiral spinor constructed like $S^\a$ but with an odd number of
$+$ signs.

In addition to mapping
the N=2 GS tensor of equation \GSfour\ onto the
N=2 RNS tensor of equation \RNStensor\ ,
this field redefinition maps the integrated GS vertex operators
(in the picture $|Z^+ Z^-|^2 \hat V$
where they have no N=2 ghost dependence) onto
integrated RNS vertex operators and maps the GS spacetime-supersymmetry
generators of eqn. \GSsusy\  onto the RNS spacetime-supersymmetry generators
$$q_\a =
 \int dz [~b\eta e^{\half(3\phi+
\int^z J_{CY}^{RNS})} S_\a
$$
$$-e^{\phi\over 2}~
\{~\psi_m\dzm x^m+ G_{CY}^{RNS}+\bar G_{CY}^{RNS} ~,~
e^{\half\int^z J_{CY}^{RNS}} S_\a~\}~],$$
$$q^*_\ad = \int dz [e^{-\half(\phi+
\int^z J_{CY}^{RNS})}S_\ad].$$

Since $R=
\int dz (\dzm \rho-\half(\ta p_\a -\tba \bar p_\ad))$
gets mapped onto the RNS
operator
$\int dz (\xi\eta-\dzm\phi)$,
the $R$-weight of a GS operator is equal to the picture of the
corresponding RNS operator. In other words, the vertex operators for
chiral fermions in the GS formalism get mapped onto Ramond states in the
$+\half$ picture, while the vertex operators for anti-chiral fermions
get mapped onto Ramond states in the $-\half$ picture. In the N=1
RNS formalism,
this identification of chirality and ghost number would be inconsistent
since it would force amplitudes to vanish unless they had a fixed number of
chiral minus anti-chiral external states. In the N=2 formalism, however,
there is no inconsistency because of the sum over instanton number which
picks up contributions from different RNS pictures.

For the six-dimensional case on $K3$, the field redefinition
relating the six-dimensional
GS variables and the conventional RNS variables
is:
$$\tilde x_{GS}^m=x_{RNS}^m,\quad
\tilde\t^\a= e^{\half(\phi+\int^z J_{K3}^{RNS})}S^\a,\quad
\tilde p_\a= e^{-\half(\phi+\int^z J_{K3}^{RNS})} S_\a,$$
$$i\dzm\tilde\sigma=cb,\quad
\dzm\tilde\rho=-2\dzm\phi+\xi\eta-J_{K3}^{RNS},$$
$$\tilde L_{K3}=L_{K3}^{RNS}+(\dzm\phi+\eta\xi)^2-
(\dzm\phi+\eta\xi)J_{K3}^{RNS},\quad
\tilde G^-_{K3}
=e^{\phi}\eta G^{-\,RNS}_{K3}$$
\eqn\sixredef{ \tilde G^+_{K3}
=e^{-\phi}\xi G_{K3}^{+\,RNS},
\quad
\tilde J_{K3}=J_{K3,RNS}+2(\dzm\phi+\eta\xi)}
where $S^\a$ is the four-component SO(5,1) chiral spinor constructed out of
the six
$\psi$ fields as $S^\a=e^{\half\int^z
(\pm \psi^0\psi^1\pm\psi^2\psi^3\pm\psi^4
\psi^5)}$ with an even number of $+$ signs, $S_\a$ is the four-component
anti-chiral spinor constructed like $S^\a$ but with an odd number of
$+$ signs, and
the GS fields with a tilde are related to those without a tilde
by the relation:
$$\tilde\Phi=\exp(-\int dz e^{i\sigma}G^+_{K3})~
\Phi~ \exp(\int dz e^{i\sigma}G^+_{K3}).
$$

As in the Calabi-Yau case, this six-dimensional field redefinition
maps the N=2 GS tensor of equation \GSsix\
into the N=2 RNS tensor of equation
\RNStensor\ and maps the GS vertex operators (in integrated form) into the
corresponding RNS vertex operators. The generator of $R$-transformations,
$\int dz [\dzm \rho+\half p_\a \t^\a]$,
is once again mapped into
the RNS
operator
$\int dz (\xi\eta-\dzm\phi)$, so $R$-weight is still equivalent to
RNS picture.

\newsec {Appendix B: The equivalence proof
of N=2 and N=3 topological amplitudes}
In this paper, we have seen that $N=1$ RNS amplitudes
are naturally phrased in the language that we have developed
for the new $N=4$ topological string, and in fact, this was a primary
motivation for defining the new theory.  In this sense,
viewing $N=2$ strings (when the matter and ghosts decouple)
as an $N=4$ topological theory is
reversing the arrow of imbedding discovered in \bv ,
where it was shown that for backgrounds where the ghost and matter
do not mix, there is a
hierarchy of imbeddings of $N=0\subset N=1\subset N=2$.
However, one may be interested in more general $N=2$
string backgrounds where the matter is mixed with the $N=2$ ghosts
(even though none is presently known, there are no reasons to doubt
their existence).
These backgrounds cannot be computed via the new topological theory
under discussion.  At any rate, it is natural to try to
generalize the hierarchy
of \bv\ to include strings with more interesting matter and ghost mixings.

A nice class of such mixed strings are the topological strings themselves.
In particular, the $N=2$ topological string may be viewed as a generalized
vacuum for the $N=0$ bosonic string where the matter and ghosts are mixed.
Similarly, $N=3$ topological strings
may be viewed as a generalized vacuum for $N=1$ superstrings where the
matter and ghosts are mixed
(see \ref\bw{M. Bershadsky, W. Lerche, D. Nemeschansky
and N. Warner, Nucl. Phys. B401 (1993) 304.}
for related discussions).  Also, one
would expect that $N=4$ topological theory could be viewed as a generalized
$N=2$ string background (see the end of this
appendix for some comments on such a background).\foot{
We should note here that there are two different
$N=4$ algebras:  the smaller one
with $SU(2)$ symmetry discussed in this paper,
and the bigger one (the `new' $N=4$ algebra)
with $SU(2)\times SU(2)$ symmetry group
which includes the flavor $SU(2)$ also in the algebra.
We have shown in this paper
that the special class of
$N=2$ string backgrounds which do not mix the matter with the ghosts
are given by the topological string based on the small $N=4$ algebra.
The more general
$N=2$ string backgrounds are presumably related to the bigger
$N=4$ algebra.}
One could therefore try to construct a topological ladder
of hierarchy, mirroring what was found in \bv :
$(N=2)_{top}\subset (N=3)_{top} \subset (N=4)_{top}^{big}$
where $(N=4)_{top}^{big}$ refers to a topological theory
based on the $SU(2)\times SU(2)$ version of the $N=4$ algebra.
In this appendix, we will explicitly
construct the first imbedding and will comment
on some difficulties in constructing the second one.

The N=0 topological partition function at genus $g$ for an $N=2$ twisted
superconformal
field theory is given by
\eqn\dpar{\prod_{j=1}^{3g-3} d^2 m_j
< | (\int\mu_j G^-)|^2>.}
where $\mu_j$ is the beltrami differential for $m_j$ and $G^-$ is the
dimension $2$ superconformal generator of the twisted N=2 theory (the other
N=2
generators are $L$ of dimension 2, $G^+$ of dimension 1, and $J$ of dimension
1).  This partition function can be understood from the
$N=0$ perspective as the vacuum amplitude
$\prod_{j=1}^{3g-3} d^2 m_j
<| (\int \mu_j b^-)|^2>$ since in the
standard technique
for getting an N=2 stress-tensor from an N=0 conformal field theory,
$G^-$ is simply $b$. Ignoring ``improvement'' terms constructed
out of U(1) currents, the other elements of the  N=2 stress-tensor
are $G^+ = J_{BRST}$
(where $Q_{N=0}=\int dz J_{BRST}$),
$J=bc$, and $L=L_{matter}+L_{ghost}$.
Note that the topological partition function of equation \dpar\
vanishes unless $\hat c=3$ ($c=9$) by U(1) charge conservation (the charge
violation must be equal to $\hat c (g-1)$).

The N=1 topological partition function at genus $g$ for an $N=3$
twisted superconformal field theory is given by
\eqn\anpar{\prod_{j=1}^{3g-3} d^2 m_j
< |(\int\mu_j
\hat G^-) \prod_{k=1}^{2g-2} \delta (\hat J^-) \hat G^3(y_k)|^2>.}
where
$(\hat G^-,\hat G^3,$
$\hat G^+)$ are the superconformal generators of dimension
$(2, 3/2,$
$ 1)$, $(\hat J^-,\hat J^3,$
$\hat J^+)$
are the SO(3) currents of dimension
$(3/2,1,$
$1/2)$,
and $\hat F$ is the fermionic generator of dimension 1/2
(the twisted N=3 tensor is related to the untwisted tensor by
$\hat L\to \hat L+\half\dzm \hat J^3, \hat G^3 \to \hat G^3 + \dzm \hat F$).
We will put hats on all $N=3$ generators in order to distinguish them
from $N=2$ generators.
The partition function of equation \anpar\
can be understood from the N=1 perspective as the
vacuum amplitude
$$\prod_{j=1}^{3g-3} d^2 m_j
< |(\int \mu_j b^-)\prod_{k=1}^{2g-2} \delta(\beta) (G_{matter}+
G_{ghost})(y_k)|^2>$$ since in the standard technique for getting
an $N=3$ stress-tensor from an $N=1$ superconformal field theory,
$\hat G^-=b$, $\hat G^3=G_{matter}+G_{ghost}$, $\hat J^-=\beta$. Ignoring
``improvement'' terms, the other elements
of the N=3 stress-tensor are $\hat L=L_{matter}+L_{ghost}$,
$\hat G^+ = J_{BRST}$, $\hat J^+ =j_{BRST}$
(where
$Q_{N=1}=\int dz J_{BRST}= \int dz \{G_{N=1}, j_{BRST}\}$),
$\hat J^3=cb-\beta\gamma$, and $\hat F=\beta c$.
Unless $\hat c=1$ ($c=3$), the topological
partition function \anpar\ vanishes by
conservation of $J^3$ charge.

Now suppose we are given an N=2 system with $c=9$. It was recently
discovered in \ref\bo{N. Berkovits and N. Ohta, {\it Embeddings
for Noncritical Superstrings}, preprint KCL-TH-94-6
and OU-HET 189, hep-th/9405144.}\
that any $(N=2,c=9)$ stress-tensor
can be embedded into an $(N=3,c=3)$
stress-tensor. It will now be shown
that the $N=1$
topological partition function of the resulting $(N=3,c=3)$ system is
equal to the $N=0$ topological partition function of the original
$(N=2,c=9)$ system, and therefore the N=0 and N=1
topological theories coming from
these two systems are equivalent.

As was shown in \bo , there exists
an embedding of any $N=2$ system with central charge $c$ into
an $N=3$ system with central charge $c-6$. This embedding does not
require a U(1) current, and when $c=6$, it reduces to an $N=3$
subset of the $N=2 \to N=4$ embedding discovered in \ref\imbe{
J. Gomis and H. Suzuki, Phys. Lett. B278 (1992) 266\semi
A. Giveon and M. Ro\v cek, Nucl. Phys. B400 (1993) 145.}.
In $N=2$ superspace notation, this embedding is given by

$$\hat T =  T -
\half\dzm(BC)+(DC)({\bar D}B)+({\bar D}C)(DB),$$
$$
 \hat G = B + C T + C(DC)({\bar D}B)+C({\bar D}C)(DB)
 - B({\bar D}C)(DC) - {c-6\over 3}[D,{\bar D}]C$$
$$+ {c-6\over 6}\sum_{n=1}^\infty (~  -C (DC)({\bar D}\dzm C)
 +C({\bar D}C)(D\dzm C) - 2(DC)({\bar D}C)[D,{\bar D}]C$$
\eqn\ttim{- \left.(2n-1)C(D{\bar D}C)({\bar D}DC)~)[ (DC)({\bar D}
C)\right]^{n-1},}
where $D= \partial_{\t}-\half\tbar\dzm$ and
${\bar D} = \partial_{\tbar}-\half\t\dzm$ are the usual $N=2$ fermionic
derivatives,
$\t_{12}\equiv
\t_1 -\t_2$, $\tbar_{12}\equiv \tbar_1 -\tbar_2$, $z_{12} \equiv
z_1 - z_2 + \half (\t_1\tbar_2 + \tbar_1\t_2)$,
$B=\tilde b+\t\beta^+ -\tbar\beta^- +\t\tbar b$
and $C=c+\t\g^+ +\tbar\g^- +\t\tbar\tilde c$
are N=2 superfields satisfying the OPE:
$$C(Z_1)B(Z_2) \to \t_{12}\tbar_{12}/z_{12},$$
$T(Z)=
J+\t G^+ +\tbar G^- + \t\tbar L$ is the untwisted
$N=2$ stress-tensor with central charge $c$ satisfying the OPE
$$T(Z_1) T(Z_2) \to$$
$$
{{c\over 3} +\t_{12}\tbar_{12} T(Z_2)\over{z_{12}^2}}
+ {{-\t_{12} DT(Z_2) +
\tbar_{12}{\bar D} T(Z_2) +
\t_{12}\tbar_{12}\dzm T(Z_2)}\over{z_{12}}},$$
and
$$(\hat T, \hat G) =(\hat J^3 + \t\hat G^+ +\tbar\hat G^-
+\t\tbar\hat L, \hat F +\t\hat J^+ + \tbar \hat J^- +\t\tbar G^3)$$
forms an untwisted
$N=3$ stress-tensor with
central charge $c-6$ satifying the OPE's:
$$\hat T(Z_1) \hat T(Z_2) \to
{{c-6\over 3} +\t_{12}\tbar_{12} \hat T(Z_2)\over{z_{12}^2}}
 + {{-\t_{12}  D\hat T +
\tbar_{12}{\bar D} \hat T(Z_2) +
\t_{12}\tbar_{12}\dzm \hat T(Z_2)}\over{z_{12}}},$$
$$\hat T(Z_1) \hat G(Z_2) \to
{{\t_{12}\tbar_{12}}\hat G(Z_2) \over 2{z_{12}^2}}
 + {{-\t_{12} D\hat G(Z_2) +\tbar_{12}{\bar D}\hat G(Z_2) +
\t_{12}\tbar_{12}\dzm \hat G(Z_2)}\over{z_{12}}},$$
$$
\hat G(Z_1) \hat G(Z_2) \to {{4c-24\over 3}+2
\t_{12}\tbar_{12}\hat T(Z_2)\over{z_{12}^2}}.$$

In a previous paper, we showed that a critical N=0 string
could be embedded into a critical N=1 string and the
non-topological N=0 and N=1 rules for computing the
scattering amplitudes gave the same result. Although
the $N=2 \to N=3$ embedding of \ttim\ is much
more complicated than the $N=0\to N=1$ embedding discussed
in \bv , the same techniques can be used to prove
equivalence of the relevant topological partition functions.
The only crucial property of the $N=2 \to N=3$ embedding is
that the ghost-number $-1$ contribution to $\hat G$ is $B$
(in components, this means that $\hat G^3=b+...$, $\hat J^+=\beta^+ +...$,
$\hat J^-=\beta^- +...$, and $\hat F=\tilde b+...$).
Furthermore note that $\hat T$ is simply the sum of $T$
and the stress-tensor for a $(B,C)$ system of
conformal weight $(1/2,-1/2)$.

After twisting $\hat T$ and $\hat G$ by sending
$\hat L\to\hat L+\half \dzm J^3$ and $\hat G^3 \to\hat G^3 +\dzm F$,
we can express the
$N=1$ topological partition function of equation \anpar\ in terms of $T$,
$B$, and $C$. Because of $BC$ conservation, the only term that contributes
to $G^3$ is $b$, the only term that contributes to $\hat G^-$ is
$G^-$, and the only term that contributes to $\delta(J^-)$
is $\delta(\beta^-)$. So the N=1 topological partition function takes the
form:
\eqn\almost{\prod_{j=1}^{3g-3} d^2 m_j
< |(\int\mu_j G^-) \prod_{k=1}^{2g-2} \delta (\beta^-) b(y_k)|^2>.}
It will now be shown that after integrating over
the $B$ and $C$ fields, this expression reduces to the
N=0 topological partition function of the original $(N=2,c=9)$ system.

The non-zero modes of these fields cancel each other out since after
twisting,
$(b,c)$ and $(\beta^-,\gamma^+)$ each have weights $(3/2,-1/2)$, while
$(\beta^+,\gamma^-)$ and $(\tilde b,\tilde c)$ each have weights $(1/2,1/2)$
(before twisting, the conformal weights of $(b,c)$ are
$(3/2,-1/2)$, of $(\beta^\pm,\gamma^\mp)$ are $(1,0)$, and of
$(\tilde b,\tilde c)$ are $(1/2,1/2)$).
The $2g-2$ zero modes of the $b$ and $\beta^-$ fields come from the $G^3$
and $\delta (J^-)$ factors in equation \anpar\ and cancel each other
out. Since there are no zero modes needed
for the other $(B,C)$ fields, expression \almost\ reduces to
$$\prod_{j=1}^{3g-3} d^2 m_j
< |(\int\mu_j G^-)|^2>,$$
which is just the N=0 topological partition function \dpar\ for the original
N=2 system.

Since this equivalence proof of the $N=1$ and $N=0$ topological
partition functions closely resembles the equivalence proof of the
$N=1$ and $N=0$ non-topological scattering amplitudes in \bv,
it is natural to ask if there is also a topological analog for
the equivalence proof of $N=2$ and $N=1$ non-topological scattering
amplitudes in \bv. In fact, there is a natural definition of
an $N=2$ topological partition function if one has a `big' twisted
$N=4$ superconformal field theory of central charge $c=0$.
This definition is
$$\prod_{j=1}^{3g-3} d^2 m_j
< |(\int\mu_j  G^-) \prod_{k=1}^{2g-2} \delta (J^-_1) G^3_1
\delta ( J^-_2) \hat G^3_2(y_k)
\prod_{m=1}^{g-1} F^-|^2>$$
where $(G^-,G^3_1,$
$G^3_2,G^+)$ are the superconformal generators
of weight $(2,3/2,$
$3/2,1)$, $(J^-_1,J^3_1,J^+_1,$
$J^-_2,J^3_2,J^+_2)$
are the $SO(4)$ currents of weight $(3/2,1,1/2,$
$3/2,1,1/2)$,
$(F^-,F^3_1,$
$F^3_2,F^+)$ are the fermionic generators of weight
$(1,1/2,$
$1/2,0)$, and $H$ is the bosonic generator of weight 0. By
conservation of $J^3_1 + J^3_2$ charge, this partition function vanishes
unless $c=0$.

Since the $N=1$ topological partition function requires an
$N=2$ superconformal field theory with $c=3$, we are therefore
looking for an embedding that takes an $(N=3,c=3)$ superconformal field
theory into an $(N=4,c=0)$ superconformal field theory.
Unfortunately,
all of the embeddings that do not require `` improvement''
terms with U(1) currents map $N=3$ stress-tensors with
central charge $c$ into $N=4$ stress-tensors with the same central charge
(this is essentially because the $N=3$ ghost system has $c=0$).
So with our present knowledge of embeddings, we are unable to find
a topological analog to the $N=1\to N=2$ non-topological embedding
of \bv .

\listrefs
\end